\documentclass[10pt,preprint]{aastex}

\begin{document}

\title{GAMER: a GPU-Accelerated Adaptive Mesh Refinement Code for Astrophysics}

\shorttitle{GAMER: a GPU-Accelerated AMR Code}

\author{Hsi-Yu Schive\altaffilmark{1,3}, Yu-Chih Tsai\altaffilmark{1}, \& Tzihong Chiueh\altaffilmark{1,2,3,4}}

\altaffiltext{1}{Department of Physics, National Taiwan University, 106, Taipei, Taiwan, R.O.C.; Email: b88202011@ntu.edu.tw (Hsi-Yu Schive)}
\altaffiltext{2}{Center for Theoretical Sciences, National Taiwan University, 106, Taipei, Taiwan, R.O.C.}
\altaffiltext{3}{Leung Center for Cosmology and Particle Astrophysics (LeCosPA), National Taiwan University, 106, Taipei, Taiwan, R.O.C.}
\altaffiltext{4}{Center for Quantum Science and Engineering (CQSE), National Taiwan University, 106, Taipei, Taiwan, R.O.C.}

\begin{abstract}
We present the newly developed code, \emph{GAMER} (\emph{GPU-accelerated Adaptive MEsh Refinement code}), which has adopted a novel approach to improve the performance of adaptive mesh refinement (AMR) astrophysical simulations by a large factor with the use of the graphic processing unit (GPU). The AMR implementation is based on a hierarchy of grid patches with an oct-tree data structure. We adopt a three-dimensional relaxing TVD scheme for the hydrodynamic solver, and a multi-level relaxation scheme for the Poisson solver. Both solvers have been implemented in GPU, by which hundreds of patches can be advanced in parallel. The computational overhead associated with the data transfer between CPU and GPU is carefully reduced by utilizing the capability of asynchronous memory copies in GPU, and the computing time of the ghost-zone values for each patch is made to diminish by overlapping it with the GPU computations. We demonstrate the accuracy of the code by performing several standard test problems in astrophysics. \emph{GAMER} is a parallel code that can be run in a multi-GPU cluster system. We measure the performance of the code by performing purely-baryonic cosmological simulations in different hardware implementations, in which detailed timing analyses provide comparison between the computations with and without GPU(s) acceleration. Maximum speed-up factors of 12.19 and 10.47 are demonstrated using 1 GPU with $4096^3$ effective resolution and 16 GPUs with $8192^3$ effective resolution, respectively.
\end{abstract}

\keywords{gravitation --- hydrodynamics --- methods: numerical}

\section{INTRODUCTION}

Numerical simulations have played an indispensable role in modern astrophysics. They serve as powerful tools to probe the fully non-linear evolutions in various problems, and provide connections between theoretical analyses and observation results. Moreover, thanks to the rapid development of the parallel computing techniques (e.g., the Beowulf clusters, Cell Broadband Engines, and graphic processing units), the spatial and mass resolutions as well as the computing performance are highly improved in the last decade.

The most essential ingredients in astrophysical simulations are the Newtonian gravity and hydrodynamics. In the last five decades, many studies have been devoted to improve both the accuracy and efficiency of numerical schemes. One of the simplest approaches is to discretize the simulation domain into a fixed number of grid cells, each of which occupies a fixed volume and position. The cell-averaged physical attributes are defined in each cell. The gravitational potential can be evaluated by several different schemes, for instance, the relaxation method, conjugate gradient method, and fast Fourier transform (FFT). As for the hydrodynamic evolution, it can be described by the modern high-resolution shock-capturing algorithms, ranging from the first-order Godunov scheme \citep{Godunov59}, the second-order monotone upwind schemes for conservation laws \citep[MUSCL;][]{vanLeer79}, to the third-order piecewise parabolic method \citep[PPM;][]{CW84}. In addition, for cosmological simulations, the particle-mesh (PM) scheme \citep[e.g.,][]{KS83,Merz05} is often adopted, in which the dark matter is treated as collisionless particles and the mass density in each grid cell is estimated by the cloud-in-cell (CIC) technique. Although this uniform-mesh method is relatively easy to implement, it suffers from the enormous memory and computing time requirements when the simulation size increases. Consequently, with the PM method, one must compromise between the size of simulation domain and the spatial resolution.

The Lagrangian particle-based approaches, in which both the collisionless dark matter and the collisional gaseous components are simulated using particles, are alternatives to the unform-mesh method. The hydrodynamic evolution is generally solved by the smoothed particle hydrodynamics \citep[SPH;][]{GM77,Lucy77}, and numerous algorithms have been adopted for solving the gravitational acceleration. The most straightforward scheme is the direct N-body method, in which all pairwise interactions are calculated. This method, while accurate, is extremely time consuming when the number of particles involved (denoted as N) is too large, owing to its ${\cal O}(N^2)$ scaling. Consequently, it is not suitable for the simulations with a large number of particles.

Several approximate algorithms have been developed to improve the computational performance of the gravitational force calculation for particle-based approaches. For example, the particle-particle/particle-mesh (${\rm P^3M}$) method \citep[e.g.,][]{HE81,Efstathiou85} improves the spatial resolution by calculating the short-range force with direct summation and adding the long-range PM force. The adaptive ${\rm P^3M}$ (${\rm AP^3M}$) method \citep[e.g.,][]{Couchman91} further improves the efficiency of force calculation by adding hierarchically refined submeshes in regions of interest and using the ${\rm P^3M}$ method locally to replace the direct pair summation. By contrast, the hierarchical tree algorithm \citep[e.g.,][]{BH86,Springel01} reduces the computational workload by utilizing the multi-pole expansion. The force exerted by distant particles is calculated using low-order multi-pole moment, and the ${\cal O}(NlogN)$ scaling is demonstrated. The Tree-PM hybrid scheme \citep[e.g.,][]{Xu95,Bagla02,Dubinski04,Springel05} serves as a further optimization to the tree algorithm. The gravitational potentials are divided into long-range and short-range terms, evaluated by the PM and tree methods, respectively. Since the tree method is applied only locally, the total workload is greatly reduced.

The particle-based approaches offer high computational performance as well as large dynamical range, owing to the Lagrangian nature. However, the main disadvantage of these approaches is their relatively poor capability for simulating hydrodynamics using the SPH method. The resolution is relatively low in high-gradient regions, for example, around the shock wave in the shock-tube test \citep{Tasker08}. It also offers poor resolution in the low-density region where the number of particles is insufficient. In addition, the SPH method suffers from the artificial viscosity, and the inability to accurately capture the hydrodynamic instabilities in certain circumstances \citep{Agertz07}.

The adaptive-mesh-refinement scheme (AMR) provides a promising approach to combine the accurate shock-capturing property of the uniform-mesh method and the high-resolution, large-dynamical-range property of the Lagrangian particle-based method. The simulation domain is first covered by uniformly distributed meshes (at the ``root'' level) with a relatively low spatial resolution, and hierarchies of nested refined meshes (at the ``refinement'' levels) with decreasing grid sizes are then allocated in regions of interest to provide the desired resolution. The gravitational potential can be computed by the multi-grid methods so that a high force resolution can be achieved in the dense region, and the grid-based shock-capturing algorithms can be applied to grids at different refinement levels to preserve the accuracy of hydrodynamic evolution. Since the simulation domain is only locally and adaptively refined, both the memory consumption and the computation time are highly reduced compared to the uniform-mesh method with the same effective resolution.

Detailed comparisons between the AMR and SPH methods have been addressed by several authors \citep[e.g.,][]{Regan07,Trac07,Tasker08,Mitchell09}. It is beyond the scope of this work. In cosmological simulations, the main drawback of the AMR method is, however, the requirement of sufficiently fine grids at the root level in order to provide adequate force resolution at the early epoch. Consequently, the memory consumption and the computation time can be larger than the Lagrangian particle-based method \citep{O'Shear05}. Nevertheless, the superior capability of handling hydrodynamic properties and the better description for low-density regions make the AMR method a promising and competitive tool in astrophysical simulations, and it has been successfully adopted for large-scale cosmological simulations \citep[e.g.,][]{Hallman09,Teyssier09}. More recently, a moving-mesh scheme has been proposed by \citet{Springel09}, aiming at integrating both the advantages of the AMR and particle-based methods.

Several approaches have been developed for the AMR implementations. The most commonly adopted approach is the block-structured AMR, which was first proposed by \citet{BO84} and \citet{BC89}. It has been implemented by many astrophysical codes, for example, Enzo \citep{BN97}, AMRA \citep{PM01}, and CHARM \citep{MC07b}. In this approach, the refined sub-domains (often referred as the mesh ``patches'') are restricted to be geometrically rectangular, and hence reduces the complexity associated with the discontinuity of resolution across different refinement levels. The size of each patch is variable and adaptable, and patches at the same refinement level can be combined or bisected to fit the local flow geometry. However, the variable patch size also leads to difficulty in parallelizing the code efficiently and to sophisticated data management. In addition, the large patch size can increase the cache-miss rate and thus lower the computational performance.

An alternative approach has been implemented in, for example, ART \citep{Kravtsov97}, MLAPM \citep{Knebe01}, and RAMSES \citep{Teyssier02}. In this approach, instead of using the rectangular patches as the basic refinement units, the refinement is performed on a cell-by-cell basis. Comparing to the block-structured AMR, it features a more efficient refinement configuration related to the local flow geometry, especially in the regions with complex geometry of structures. However, the main drawback of this method lies in the more sophisticated data management, owing to its irregular shape of domain refinement. The interface profiles between cells with different zone spacings are complex, and spatial interpolations must be frequently used to provide the boundary conditions for each cell. Moreover, since the size of stencils required by the hydrodynamic and Poisson solvers for each cell is generally much larger than a single cell, the computational overhead is large and can lead to serious performance deterioration.

The FLASH code \citep{Fryxell00}, which uses the PARAMESH AMR library \citep{MacNeice00}, and the NIRVANA code \citep{Ziegler05} have adopted a third approach for the AMR implementation. In this approach, the domain refinement is based on a hierarchy of mesh patches similar to the block-structured AMR, whereas each patch is restricted to have the same number of cells. The typically adopted patch sizes are $8^3$ in FLASH and $4^3$ in NIRVANA. Although this restriction will certainly impose the inflexibility of domain refinement and result in a relatively large refined volume compared to the two methods described above, it features several important advantages. First, the data structure and the interfaces between neighboring patches are considerably simplified, which can lead to significant improvements of performance and parallel efficiency. Second, since the additional buffer zones (often referred as the ''ghost zones'') for finite-difference stencils are only needed for each patch instead of each cell, the computational overhead associated with the preparation of the ghost-zone data is greatly reduced compared to the cell-based refinement strategy. Finally, fixing the patch size allows for easier optimization of performance and also increases the cache-hit rate. All these features are essential for developing a high-performance code, especially for parallel computing such as using graphic processing units. Accordingly, in \emph{GAMER}, we have adopted this approach as the refinement strategy.

Novel use of modern graphic processing units (GPU) for acceleration of numerical calculations has becoming a widely-adopted technique in the past three years. The original purpose of GPU is to serve as an accelerator for computer graphics. It is designed to work with the Single Instruction, Multiple Data (SIMD) architecture, and processes multiple vertex and fragment data in parallel. The modern GPU, for example the NVIDIA Tesla C1060, has 240 scalar processor cores working at 1.3 GHz clock rate. It delivers a peak performance of 933 GFLOPS (Giga Floating Operations per Second), which is about an order of magnitude higher than the modern CPU. In addition, it has 4 GB GDDR3 internal memory with memory bandwidth of 102 GB/s. The 240 scalar processor cores are grouped into 30 multiprocessors, each of which consists of 8 scalar processor cores and shares a 16 KB on-chip data cache. The NVIDIA Tesla S1070 computing system further combines four Tesla C1060 GPUs and offers a nearly 4 TFLOPS computing power. Given the natural capability of parallel computing and the enormous computational power of GPU, using GPU for general-purpose computations (GPGPU\footnote{http://gpgpu.org/}) have become an active area of research.

The traditional scheme in GPGPU works by using the high-level shading languages, which are designed for graphic rendering and require familiarity with computer graphics. It is therefore difficult and unsuitable for general-purpose computations. In 2006, the NVIDIA Corporation releases a new computing architecture in GPU, the Compute Unified Device Architecture \citep[CUDA;][]{NVIDIA08}. It is designed for a general-purpose usage and greatly lowers the threshold of using GPU for \emph{non-graphic} computations. In CUDA, GPU is regarded as a multi-threaded coprocessor to CPU with a standard C language interface. To define the computational task for GPU, programmers should provide a C-like function called ``kernel'', which can be executed by multiple ``CUDA threads'' in parallel. An unique thread ID is given to each thread in order to distinguish between different threads.

As an illustration, we consider the sum of two vectors, each of which has M elements. In CUDA, instead of writing a loop to perform M summation operations sequentially, we define a single summation operation in a kernel and use M threads. These M threads will execute the same kernel in parallel but perform the single summation operation on different vector elements. In this example, the thread ID may be used to defined the targeted vector element for each thread.

Note that this scenario is analogous to the parallel computing in a Beowulf cluster using the message passing interface (MPI), in which a single program is simultaneously executed by multiple processors and each process is given an unique ID (``MPI rank''). However, performance optimization in GPU is not straightforward and requires elaborate numerical algorithms dedicated to the GPU specifications. Especially, note that CUDA threads are further grouped into multiple ``thread blocks''. Threads belonging to the same thread block are executed by one multiprocessor and can share data through an on-chip data cache (referred as the ``shared memory''). This data cache is very small (typically only 16 KB per multiprocessor) but has much lower memory latency than the off-chip DRAMS (referred as the ``global memory''). Accordingly, the numerical algorithms must be carefully designed to store common and frequently used data in this fast shared memory so that the memory bandwidth bottleneck may be removed.

Nowadays, the most successful approach to utilize the GPU computing power in astrophysical simulations is the direct N-body calculation \citep[e.g.,][]{Belleman08,Schive08,Gaburov09}. \citet{Schive08} have built a multi-GPU computing cluster named GraCCA (Graphic-Card Cluster for Astrophysics), and have demonstrated its capability for the direct N-body simulations in terms of both the high computational performance as well as the high parallel efficiency. The direct calculation of all $N^2$ pairwise interactions is extremely computational-intensive, and thus is relatively straightforward to obtain high performance in GPU. However, the direct N-body simulations can address only a limited range of problems. \citet{Aubert09} have proposed a GPU-accelerated PM integrator using a single GPU. It remains considerably challenging and unclear whether the performance of other kinds of astrophysical simulations with complex data structure and relatively low arithmetic intensity, such as the AMR simulations, can be highly improved by using GPUs, especially in a multi-GPU system.

In this paper, we present the first GPU-accelerated, adaptive mesh refinement, astrophysics-dedicated, and parallelized code, named \emph{GAMER} (\emph{GPU-accelerated Adaptive MEsh Refinement code}). We give a detailed description of the numerical algorithms adopted in this code, especially focusing on the GPU implementations. The accuracy of the code is demonstrated by performing various test problems. Detailed timing analyses of individual GPU solvers as well as the complete program are conducted with different hardware implementations. In each timing test, we further compare the performances of runs with and without GPU(s) acceleration.

The paper is organized as follows. In \S\ 2 we describe the numerical schemes adopted in \emph{GAMER}, including the AMR method, both the hydrodynamic scheme and the Poisson solver, and the parallelization strategy. We then focus on the GPU implementations of different parts in the code, along with individual performance measurements in \S\ 3. In \S\ 4, we present the simulation results of several test problems to demonstrate the accuracy. Detailed timing analyses of the complete program in purely-baryonic cosmological simulations are presented in \S\ 5. Finally, we summarize the work and discuss the future outlooks in \S\ 6.

\section{NUMERICAL SCHEME}

In this section, we describe in detail the numerical schemes adopted in \emph{GAMER}, including the AMR implementation, the algorithms of both hydrodynamics and self gravity, and the parallelization strategy. To provide a more comprehensible description, here we focus on the generic algorithms that are unrelated to the hardware implementation. Important features related to the GPU implementation will be emphasized and a more detailed description will be given in \S\ 3.

\subsection{Adaptive Mesh Refinement}

\begin{figure}
\centering
\includegraphics[width=10cm]{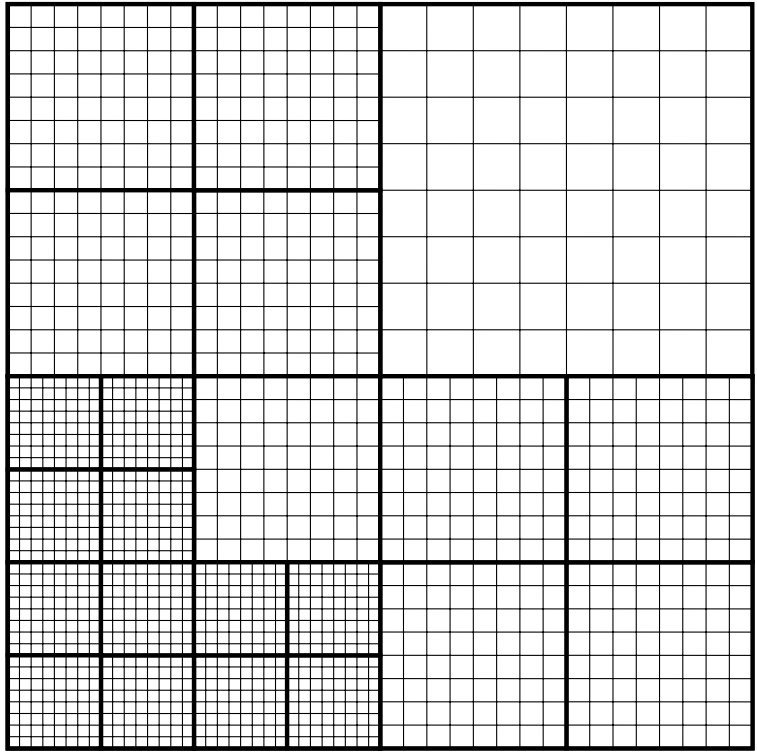}
\caption{Two-dimensional example of the refinement map. Each patch is composed of $8\times8$ cells. The borders of patches are highlighted by bold lines. One patch may be refined into four child patches with a spatial resolution twice that of their parent patch. A jump of more than one refinement level between nearby patches is forbidden (even in the diagonal direction).}
\label{fig:RefinementMap}
\end{figure}

The AMR scheme implemented in \emph{GAMER} is similar to that adopted by FLASH, in which the computational domain is covered by a hierarchy of grid patches with similar shape but different spatial resolutions. In \emph{GAMER}, a grid patch is defined to have a fixed number of grid cells in each spatial direction. The computational domain is first covered by root patches with the lowest spatial resolution. Then, according to the user-defined refinement criteria, each root patch may be refined into eight child patches with a spatial resolution twice that of their parent patch. The same refinement operation may be further applied to all patches in different refinement levels, in which a patch at level $\ell$ has a spatial resolution $2^\ell$ times higher than that of a root patch at level zero. Accordingly, a hierarchy of grid patches with oct-tree data structure is dynamically and adaptively constructed during the simulation. In Figure \ref{fig:RefinementMap}, we show a two-dimensional example of the refinement map.

Patches are the basic units in \emph{GAMER}. Owing to the oct-tree data structure, eight patches are always allocated or deallocated simultaneously. The data stored in each patch include its own physical variables, the absolute coordinates in the computational domain, the indices of the parent, child, and 26 sibling patches, the pointers of flux arrays corresponding to 6 patch boundary surfaces, and the flag recording its refinement status.

Restricting all patches to be geometrically similar to each other greatly simplify the AMR framework, with respect to both the structure of the program as well as the GPU implementation. A single GPU kernel can be applied to all patches, even in different refinement levels. Moreover, since the amount of computation workload of each patch is the same, there will be no synchronization overhead when multiple patches are evolved in parallel by GPU. However, it does impose certain inflexibility of spatial refinement. The region being refined will be larger than necessary, especially when the volume of a single patch is too large. On the other hand, having a small-volume patch will introduce higher computational overhead associated with the preparation of the ghost-zone data. In \emph{GAMER}, the optimized size of a single patch is set to $8^3$. It will be demonstrated in \S\ 3 and \S\ 5 that by exploiting the feature of parallel execution between CPU and GPU, the ghost-zone filling time can be overlapped with the execution time of the GPU solvers, and yields considerable performance enhancement.

\emph{GAMER} can be used as either a purely hydrodynamic or a coupled self-gravity and hydrodynamic code. When only the hydrodynamic module is activated, the code supports both the uniform and individual time step algorithms. The time step of level $\ell$ may be either equal to or twice smaller than that of level $\ell-1$. However, when the gravity module is also activated, the code currently only supports the uniform time step algorithm. The same time step is applied to all levels, and the evolution of patches at level $\ell$ proceeds in the steps as follows.
\begin{enumerate}
\item Update physical quantities for all patches at level $\ell$.
\item Begin the evolution of the next refinement level if there are patches at level $\ell+1$.
\item Correct the physical quantities at level $\ell$ by using the updated results at level $\ell+1$.
\item Rebuild the refinement map at level $\ell$.
\end{enumerate}

Since the fine-grid values are presumably more accurate than the coarse-grid values, there are two cases where the data of a coarse patch require further correction in the step 3. First, if a coarse patch is overlaid by its child patches, its values are simply replaced by the spatial average of the fine-grid values. Second, if the border of a coarse patch is near the boundary of refinement, the flux correction operation \citep{BC89} is applied. First, a corresponding flux array is allocated. This array will store the difference between the coarse-grid flux and the fine-grid flux across the coarse-fine boundary, and will be used to correct the coarse-grid values adjacent to this boundary. This flux correction operation ensures that the flux out of the coarse-grid patch is equal to the flux into the fine-grid patch, and therefore the conservation of hydrodynamic variables is preserved (assuming no self gravity).

Rebuilding the refinement map in the step 4 takes two sub-steps: first a flag step, followed by a refinement step. A patch is flagged for refinement if any cell inside the patch satisfies the refinement criteria. In \emph{GAMER}, both the hydrodynamic variables and their gradients can be taken as the refinement criteria. There is however a situation requiring special treatment during the flag step. Since the refinement map is always rebuilt from finer levels to coarser levels, a patch at level $\ell$ may not be flagged even if its child patches at level $\ell+1$ have already been flagged. In this case, the patch at level $\ell$ is also flagged to ensure that the fine-grid data are preserved. Finally, a proper-nesting constraint is applied to all patches. It prohibits the spatial refinement from jumping more than one level across two adjacent patches. Patches failing to satisfy this constraint are unflagged.

In the refinement step, eight child patches at level $\ell+1$ are constructed for each flagged patch at level $\ell$. The hydrodynamic data of child patches are either directly inherited from existing data or filled via conservation-preserving interpolation from their parent patches. The Min-Mod limiter is used to ensure the monotonicity of interpolation. The indices of parent, child, and sibling patches are stored, and null values are assigned to them if the corresponding child or sibling patches do not exist. Finally, the flux arrays are properly allocated for patches adjacent to the coarse-fine boundaries.

\begin{figure}
\centering
\includegraphics[width=10cm]{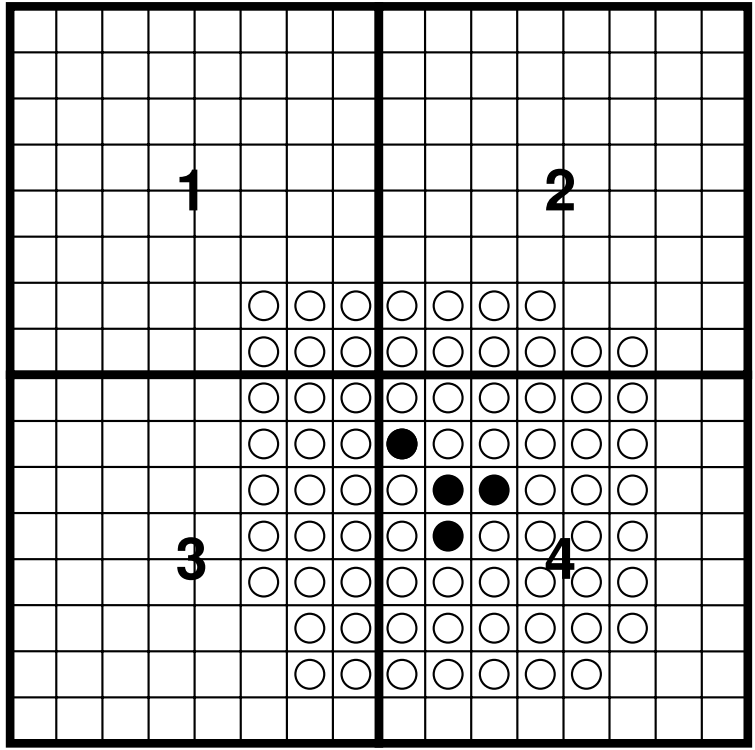}
\caption{Two-dimensional example of the flag operation. The size of the flag buffer ($N_b$) is set to 3. The borders of the patches are highlighted by bold lines. The filled circles represent the cells satisfying the refinement criteria, and the open circles represent their corresponding flag buffers. The numbers at the center of each patch stand for the patch indices. In this example, although there are no cells fulfilling the refinement criteria in the patches 1, 2, and 3, they are stilled flagged due to the extension of the flag buffers.}
\label{fig:FlagBuffer}
\end{figure}

The frequency of rebuilding the refinement map is also a free parameter provided by users. The guideline is that the refinement map must be rebuilt before the regions of interest propagate away from fine-grid patches into coarse-grid patches. Although in the extreme case we may rebuild the refinement map in every step, it is too expensive in time and not practical in general situations. Therefore, in order to reduce the frequency of performing the refinement operation, we follow the scheme suggested by \citet{BC89}. A free parameter $N_b$ is provided to define the size of the \emph{flag buffer}. If a cell exceeds the refinement threshold during the flag check, $(1+2N_b)^3-1$ cells surrounding this cell are regarded as the flag buffers. If any of these flag buffers extends across the patch border, the corresponding sibling patch is also flagged for refinement (as long as it satisfies the proper-nesting condition). Figure \ref{fig:FlagBuffer} shows an example of the refinement result with $N_b=3$. The extreme case is to have $N_b$ equal to the size of a single patch, in which case all 26 sibling patches will always be flagged if the central patch is flagged. Generally speaking, the larger the number $N_b$, the longer period between two refinement steps may be adopted.

The procedure of patch construction during the initialization is different from that during the simulation. As illustrated by the evolution procedure described previously, the patch construction during the simulation is always performed from finer levels to coarser levels. It ensures that the fine-grid data are predominant of patch refinement. By contrast, the spatial resolution of the initial condition is solely provided by users. Accordingly, in \emph{GAMER}, three kinds of initialization methods are supported.

First, a user-defined initialization function can be applied to set the initial value of each physical quantity. The patch construction starts from level 0 up to the maximum level. If any patch at level $\ell$ satisfies the refinement criteria, eight child patches at level $\ell+1$ are allocated and initialized by the same initialization function. After patch construction, a restriction operation is performed, starting from the maximum level down to the root level, in order to ensure that the physical quantities of a cell at level $\ell$ is always equal to the spatial average of its eight child cells at level $\ell+1$.

Second, the code can load an array storing the uniform-mesh data as the initial condition. Assuming that the input data have spatial resolution equal to the refinement level $\ell$, then patches at level $\ell$ are first constructed. After that, a restriction operation is performed from level $\ell$ down to level 0 to construct patches of levels $< \ell$. Any patch failing to satisfy the refinement criteria at the current level will be removed. Since we assume that the input uniform-mesh data possess the highest resolution during the initialization, no patch at level $> \ell$ is allocated at this stage. However, patches of higher resolution can still be constructed during the run, and hence the highest resolution is not limited by the initial input data.

The third initialization procedure loads any of the previous data dumps as the restart file. It is essential when the program is terminated unexpectedly, and it also provides an efficient way for tuning parameters and analyzing simulation results.

\subsection{Hydrodynamics}
In \emph{GAMER}, the Euler equations are solved in conservative forms:
\begin{equation}
\frac{\partial\rho}{\partial t}+\frac{\partial}{\partial x_j}(\rho v_j)=0,
\label{eq:MassConserve}
\end{equation}
\begin{equation}
\frac{\partial(\rho v_i)}{\partial t}+\frac{\partial}{\partial x_j}
(\rho v_iv_j+P\delta_{ij})=-\rho\frac{\partial\phi}{\partial x_i},
\label{eq:MomConserve}
\end{equation}
\begin{equation}
\frac{\partial e}{\partial t}+\frac{\partial}{\partial x_j}[(e+P)v_j]=-\rho v_j\frac{\partial\phi}{\partial x_j},
\label{eq:EnergyConserve}
\end{equation}
where $\rho$ is the mass density, $v$ is the flow velocity, $P$ is the thermal pressure, $e$ is the total energy density, and $\phi$ is the gravitational potential. The relation between pressure $P$ and total energy density $e$ is given by
\begin{equation}
e=\frac{1}{2}\rho v^2+\epsilon,
\label{eq:ThermalEnergy}
\end{equation}
\begin{equation}
P=(\gamma-1)\epsilon,
\label{eq:EOS}
\end{equation}
where $\epsilon$ is the internal thermal energy density and $\gamma$ is the ratio of specific heats. The self gravity is included in the Euler equations as a source term, and will be addressed in more detail in the next subsection.

The hydrodynamic scheme adopted in \emph{GAMER} is based on the algorithm proposed by \citet{TP03}. It is a second-order accurate relaxing total variation diminishing (TVD) scheme \citep{JX95}, which has been implemented and well tested in both the hydrodynamic simulation \citep{TP03} as well as the magnetohydrodynamic simulation \citep{Pen03}. In the following, we first review the one-dimensional relaxing TVD scheme, and then follow the generalization to the three-dimensional case.

Consider the one-dimensional Euler equation in vector form:
\begin{equation}
\frac{\partial {\bf u}}{\partial t}+\frac{\partial {\bf F}({\bf u})}{\partial x}=0,
\label{eq:1DEuler}
\end{equation}
where ${\bf u}=(\rho,\rho v,e)$ is the flow-variable vector and ${\bf F}({\bf u})=(\rho v,\rho v^2+P,ev+Pv)$ is the corresponding flux vector. First, a free positive function $c(x,t)$, which is referred as the freezing speed, is evaluated and an auxiliary vector is defined by ${\bf w}\equiv{\bf F}({\bf u})/c$. To guarantee the TVD condition, the freezing speed $c$ must be greater than the speed of information propagation. For the one-dimensional Euler equation, this requirement is satisfied by having $c(x,t)=|v(x,t)|+c_s(x,t)$, where $c_s$ is the sound speed.

The flux term in Eq. (\ref{eq:1DEuler}) is then decomposed into two terms,
\begin{equation}
\label{eq:1DEulerLRFlux}
\frac{\partial {\bf u}}{\partial t}+\frac{\partial {\bf F}^R}{\partial x}-\frac{\partial {\bf F}^L}{\partial x}=0,
\end{equation}
where
\begin{equation}
{\bf F}^R\equiv c\left(\frac{{\bf u}+{\bf w}}{2}\right) ~~ {\rm and} ~~
{\bf F}^L\equiv c\left(\frac{{\bf u}-{\bf w}}{2}\right)
\label{eq:LRFlux}
\end{equation}
are referred as the right-moving and left-moving fluxes with advection speed $c$, respectively. Since these two fluxes have well-defined directions, the MUSCL scheme can be applied straightforwardly. Let ${\bf u}_n^t$ denote the cell-centered value of the cell $n$ at time $t$, and ${\bf F}_n^t$ denote the corresponding cell-center flux. To integrate Eq. (\ref{eq:1DEulerLRFlux}) in a conservative form, the fluxes ${\bf F}_{n\pm 1/2}^{R,t}$ and ${\bf F}_{n\pm 1/2}^{L,t}$ defined at the boundaries of the cell $n$ must be evaluated. In the following, we describe the algorithm to evaluate ${\bf F}_{n+1/2}^{R,t}$ as an illustration. ${\bf F}_{n-1/2}^{R,t}$ and ${\bf F}_{n\pm 1/2}^{L,t}$ can be derived in a similar way.

In the first step, the upwind scheme is used to assign value to the boundary flux as a first-order approximation. Since ${\bf F}_{n+1/2}^{R,t}$ has a positive advection velocity, we can simply set ${\bf F}_{n+1/2}^{R,t}={\bf F}_n^t$. The second-order correction $\triangle {\bf F}_{n+1/2}^{{\rm TVD},t}$ satisfying the TVD condition is obtained by applying a flux limiter $\phi$ to two second-order flux corrections,
\begin{equation}
\triangle {\bf F}_{n+1/2}^{{\rm TVD},t}=\phi(\triangle {\bf F}_{n+1/2}^{(1),t},\triangle {\bf F}_{n+1/2}^{(2),t}),
\label{eq:2ndTVDFluxCorr}
\end{equation}
where
\begin{equation}
\triangle {\bf F}_{n+1/2}^{(1),t}=\frac{{\bf F}_n^t-{\bf F}_{n-1}^t}{2} ~~ {\rm and} ~~
\triangle {\bf F}_{n+1/2}^{(2),t}=\frac{{\bf F}_{n+1}^t-{\bf F}_n^t}{2}.
\label{eq:2ndFluxCorr}
\end{equation}
The flux limiter adopted in the current implementation is the van Leer limiter \citep{vanLeer74}, which takes the harmonic average of two second-order flux corrections:
\begin{equation}
\phi_{vanLeer}(\triangle {\bf F}^{(1)},\triangle {\bf F}^{(2)})=\left\{
\begin{array}{ll}
\frac{2\triangle {\bf F}^{(1)}\triangle {\bf F}^{(2)}}{\triangle {\bf F}^{(1)}+\triangle {\bf F}^{(2)}},
~~~ {\rm if} ~~ \triangle {\bf F}^{(1)}\triangle {\bf F}^{(2)}>0,\\
\\
0, ~~~~~~~~~~~~~~~~~ {\rm if} ~~ \triangle {\bf F}^{(1)}\triangle {\bf F}^{(2)}\le 0.
\end{array}
\right.
\label{eq:vanLeer}
\end{equation}
Note that, as indicated by Eq. (\ref{eq:vanLeer}), no second-order correction is applied to ${\bf F}_{n+1/2}^{R,t}$ if ${\bf F}_n^t$ assumes a local extreme value, and hence the hydrodynamic scheme is locally reduced to only first-order accurate.

Finally, the second-order accurate right-moving flux is given by
\begin{equation}
{\bf F}_{n+1/2}^{R,t}={\bf F}_n^t+\triangle {\bf F}_{n+1/2}^{{\rm TVD},t}.
\label{eq:2ndRFlux}
\end{equation}
${\bf F}_{n-1/2}^{R,t}$ and ${\bf F}_{n\pm 1/2}^{L,t}$ can be evaluated in the way similar to Eq. (\ref{eq:2ndRFlux}).

To achieve second-order accuracy in time as well, the second-order Runge-Kutta method (also known as the midpoint method) is adopted for the time integration. First, the temporal midpoint value ${\bf u}^{t+\triangle t/2}_n$ is evaluated by
\begin{equation}
{\bf u}^{t+\triangle t/2}_n={\bf u}^t_n-\left(\frac{{\bf F}_{n+1/2}^t-{\bf F}_{n-1/2}^t}{\triangle x}\right)
\frac{\triangle t}{2},
\label{eq:MidpointValue}
\end{equation}
where ${\bf F}_{n+1/2}^t={\bf F}_{n+1/2}^{R,t}-{\bf F}_{n+1/2}^{L,t}$ is computed by the first-order upwind scheme. The midpoint fluxes ${\bf F}^{t+\triangle t/2}$ are then computed by applying the second-order TVD scheme to ${\bf u}^{t+\triangle t/2}$. Eventually, the full-step value ${\bf u}_n^{t+\triangle t}$ is given by
\begin{equation}
{\bf u}^{t+\triangle t}_n={\bf u}^t_n-\left(\frac{{\bf F}_{n+1/2}^{t+\triangle t/2}
-{\bf F}_{n-1/2}^{t+\triangle t/2}}{\triangle x}\right)\triangle t.
\label{eq:FullstepValue}
\end{equation}

It is straightforward to generalize the one-dimensional TVD scheme described above to three dimensions by applying the dimensional splitting method \citep{Strang68}. The three-dimensional Euler equations are solved by first applying a forward sweep in the order $xyz$, and follows a backward sweep in the order $zyx$. The same time step must be employed by these two sweeps to maintain the second-order accuracy. The dimensional spitting method also makes it easy to parallelize the computation of the three-dimensional Euler equations, as addressed by \citet{TP03}. By taking advantage of this feature, a high-performance GPU hydrodynamic solver based on the above TVD scheme has been implemented in \emph{GAMER}. It will be described in detail in \S\ 3.1.

The one-dimensional TVD scheme uses a seven-point stencil (one cell on each side for evaluating the midpoint values by the upwind scheme plus two cells on each side for evaluating the full-step values by the TVD scheme). Therefore, three ghost zones are required on each side in each spatial direction to update the hydrodynamic variables in a single patch. The ghost-zone values are filled in two ways. If a desired sibling patch exists, they are filled by a direct memory copy. Otherwise they are filled by linear interpolation with the Min-Mod limiter from patches one level coarser. Since the proper-nesting condition is fulfilled everywhere in the simulation domain, interpolation from patches two (or more) levels coarser is prevented. Note that computing the ghost-zone values can lead to a significant computational overhead in almost all kinds of AMR implementations. Nevertheless, this issue is well handled in \emph{GAMER} and will be addressed in \S\ 3 and \S\ 5.

\emph{GAMER} can work in both the physical coordinates as well as the comoving coordinates. For the cosmological hydrodynamic simulation, the forms of Euler equations (Eqs. [\ref{eq:MassConserve}]-[\ref{eq:EnergyConserve}]) may remain unchanged by applying the following changes of variables:
\begin{equation}
\tilde{{\bf x}}= \frac{{\bf x}}{a}, ~~~~~~ d\tilde{t}=\frac{dt}{a^2}, ~~~~~~ \tilde{\rho}=a^3\rho,
\label{eq:ChangeOfVariable1}
\end{equation}
\begin{equation}
\tilde{{\bf v}}=a({\bf v}-H{\bf x}), ~~~~~~ \tilde{P}=a^5P, ~~~~~~ \tilde{\phi}=a^2(\phi-\phi_b),
\label{eq:ChangeOfVariable2}
\end{equation}
where $a$ is the cosmological scale factor, $H$ is the Hubble parameter, and $\phi_b$ is the gravitational potential related to the background density (here we have assumed that the ratio of specific heats $\gamma=5/3$). It makes the \emph{GAMER} code more flexible and hence can be applied to different aspects of astrophysical applications.

\subsection{Self Gravity}
The gravitational potential is evaluated via solving the Poisson equation
\begin{equation}
\nabla^2\phi({\bf x})=4\pi G\rho({\bf x}),
\label{eq:Poisson}
\end{equation}
where $G$ is the gravitational constant. In its discrete form, the Laplacian operator $\nabla^2$ can be replaced by a 7-points finite difference operator:
\begin{equation}
\frac{1}{\triangle h_\ell^2}\left(\phi_{i+1,j,k}^{t,\ell}+\phi_{i,j+1,k}^{t,\ell}+\phi_{i,j,k+1}^{t,\ell}
+\phi_{i-1,j,k}^{t,\ell}+\phi_{i,j-1,k}^{t,\ell}+\phi_{i,j,k-1}^{t,\ell}-6\phi_{i,j,k}^{t,\ell}\right)
=4\pi G\rho_{i,j,k}^{t,\ell},
\label{eq:DiscretePoisson}
\end{equation}
where $\triangle h_\ell$ is the zone spacing at level $\ell$.

In \emph{GAMER}, two numerical methods have been implemented to solve Eq. (\ref{eq:DiscretePoisson}). At the root level, where the coarsest patches always cover the entire computational domain, we adopt the standard fast Fourier transform (FFT) method. A Green's function associated with the discretized Laplacian operator is used, and the periodic boundary condition is assumed. At the refined levels, where in general the computational domain is only partially refined and hence Eq. (\ref{eq:DiscretePoisson}) cannot be solved globally, we adopt the successive overrelaxation method \citep[SOR;][]{Press07} with the Dirichlet boundary condition. Only one ghost zone is required on each side in each spatial direction to evaluate the potential in a single patch, and the ghost-zone values are filled by interpolation from the patches one level coarser.

The SOR scheme starts with evaluating the residual of each cell by
\begin{equation}
R_{i,j,k}^\ell=(\phi_{i+1,j,k}^{old,\ell}+\phi_{i,j+1,k}^{old,\ell}+\phi_{i,j,k+1}^{old,\ell}
+\phi_{i-1,j,k}^{old,\ell}+\phi_{i,j-1,k}^{old,\ell}+\phi_{i,j,k-1}^{old,\ell}-6\phi_{i,j,k}^{old,\ell})
-4\pi G\rho_{i,j,k}^\ell\triangle h_\ell^2,
\label{eq:Residual}
\end{equation}
where the superscript $old$ indicates the values at a previous step. The updated values are then given by
\begin{equation}
\phi_{i,j,k}^{new,\ell}=\phi_{i,j,k}^{old,\ell}+\frac{1}{6}\omega R_{i,j,k}^\ell,
\label{eq:SOR}
\end{equation}
where $\omega$ is the overrelaxation parameter. Eqs. (\ref{eq:Residual}) and (\ref{eq:SOR}) are solved iteratively until the one-norm of the residual has been diminished to the floating-point precision limit (compared to the source term $4\pi G\rho\triangle h^2$) to ensure the solution of potential has converged.

The odd-even ordering is adopted to determine the order in which different cells in a patch are updated by Eqs. (\ref{eq:Residual}) and (\ref{eq:SOR}). A cell with position indices $(i,j,k)$ is regarded as an ``even'' cell if $i+j+k$ is even, and ``odd'' cell if $i+j+k$ is odd. This scheme is also referred as the ``red-black ordering'' since the odd and even cells are defined in the way just like the red and black squares of a checkerboard in the two-dimensional case. During one iteration, the even cells are first updated, and then these \emph{updated} values are used to update the odd cells. As can be seen from Eq. (\ref{eq:Residual}), the updates of even cells depend only on the odd cells, and vice versa. Therefore, by exploiting the odd-even ordering, all even cells can be calculated independently at the first half-step (hereafter referred as the \emph{even step}), and all odd cells can also be calculated independently at the second half-step (hereafter referred as the \emph{odd step}). The independent operations reveal the possibility of parallel computing. This property plays a crucial role in the development of an efficient and memory-saving GPU kernel for the SOR scheme.

In real astrophysical simulations, generally the number of root-level patches takes only a small fraction ($<10\%$) of the total number of patches in all levels, and hence the execution time of the root level is much less than the total simulation time. Accordingly, for the root-level Poisson solver, we execute on the CPU the free available package FFTW \citep{FJ98}, which is a highly-optimized, parallelized, and portable package for solving the discrete Fourier transform (DFT).  On the contrary, for the refinement levels where the Poisson equation is solved via the SOR scheme, a high-performance GPU Poisson solver has been implemented in \emph{GAMER}. It will be described in detail in \S\ 3.2.

Solving the Poisson equation in an AMR framework requires additional attention. The primary issue is that the boundary condition for a fine-grid patch is always obtained by interpolation from the coarse-grid values. The potential is continuous across the coarse-fine interface; however, the normal derivative of potential is not necessarily so. The discontinuity in normal derivative of potential acts as a \emph{pseudo mass sheet} on the coarse-fine interface, and will eventually contaminate the solution of potential in finer levels.

\begin{figure}
\centering
\includegraphics[width=10cm]{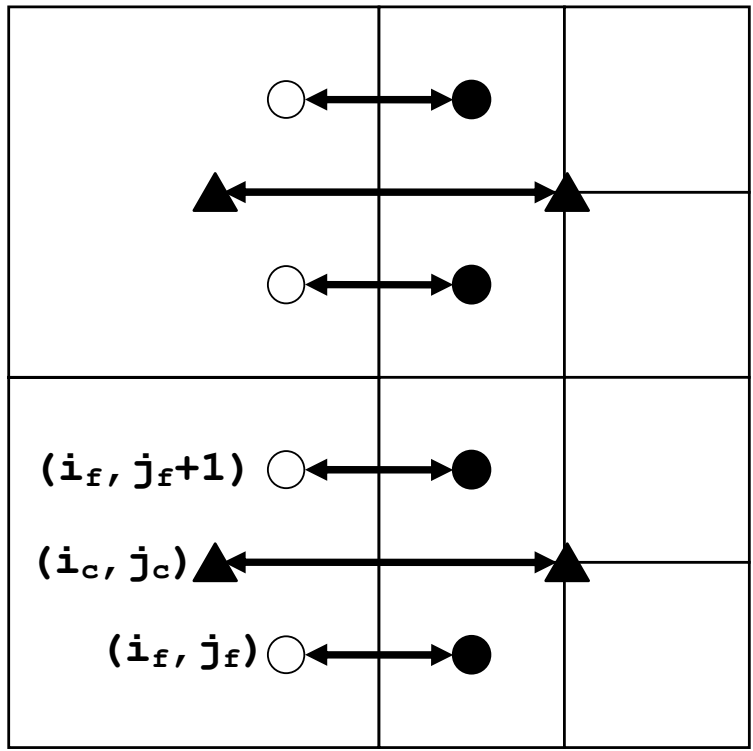}
\caption{Two-dimensional example of the mesh structure adjacent to a coarse-fine interface. The filled triangles represent the coarse-grid potentials, the filled circles represent the fine-grid potentials, and the open circles represent the fine-grid potentials in the ghost zones. The values used for approximating the normal derivatives of potentials are connected by two-way arrows.}
\label{fig:GetPseudoRho}
\end{figure}

Several methods have been proposed to improve the solution of the Poisson equation in adaptively refined meshes \citep[e.g.,][]{MC96,HG00,Ricker08}. In \emph{GAMER}, the two-level potential correction is in work with the following procedure. First, we estimate the pseudo mass sheet on all interfaces between a parent patch and each of its eight child patches. A two-dimensional example of the mesh structure adjacent to a coarse-fine interface is shown in Figure \ref{fig:GetPseudoRho}, in which $\phi_{i_c,j_c}^c$ denotes a coarse-grid potential to the left of the interface, and $\phi_{i_f,j_f}^f$ and $\phi_{i_f,j_f+1}^f$ denote the corresponding two fine-grid potentials (defined in the ghost zones). The pseudo mass sheet $\xi$ is then defined by
\begin{equation}
\xi({\bf x})=\frac{1}{4\pi G\triangle h^c}\left(\left.\frac{\partial\phi}{\partial n}\right|_{\epsilon^-}-\left.\frac{\partial\phi}{\partial n}\right|_{\epsilon^+}\right),
\label{eq:PseudoRho}
\end{equation}
in which the normal derivatives of potentials are approximated by
\begin{equation}
\left.\frac{\partial\phi}{\partial n}\right|_{\epsilon^-}=\frac{\phi_{i_c,j_c}^c-\phi_{i_c+1,j_c}^c}{\triangle h^c}
\label{eq:CoarseGradientPhi}
\end{equation}
on the coarse-grid side and
\begin{equation}
\left.\frac{\partial\phi}{\partial n}\right|_{\epsilon^+}=\frac{(\phi_{i_f,j_f}^f+\phi_{i_f,j_f+1}^f
-\phi_{i_f+1,j_f}^f-\phi_{i_f+1,j_f+1}^f)/2}{\triangle h^f}
\label{eq:FineGradientPhi}
\end{equation}
on the fine-grid side.

To compensate the mass-sheet potentials, we first place the \emph{negative} of the pseudo mass sheet in the coarse meshes adjacent to the coarse-fine boundary, and evaluate the correction to the coarse-grid potential (denoted as $\zeta^c$) by solving the correction equation
\begin{equation}
\nabla^2\zeta({\bf x})=-4\pi G\xi({\bf x}).
\label{eq:PotCorrection}
\end{equation}
Afterwards, the fine-grid correction can also be solved by Eq. (\ref{eq:PotCorrection}), in which the boundary condition is provided through the interpolation of the coarse-grid correction. Finally, the corrected potentials in both coarse and fine grids are obtained by
\begin{equation}
\phi_{corrected}({\bf x})=\phi_{uncorrected}({\bf x})+\zeta({\bf x}).
\label{eq:CorrectedPot}
\end{equation}

Note that for AMR implementations that demand patches in every refinement level to have identical shape (e.g., the FLASH and \emph{GAMER} codes), the Poisson equation is solved on a patch-by-patch basis in order to diminish the data transfer between adjacent patches \citep{Ricker08}. Consequently, similar numerical errors are also introduced from the interfaces between patches at the same refinement level. Nevertheless, since Eqs. (\ref{eq:PseudoRho})-(\ref{eq:FineGradientPhi}) are used at \emph{all interfaces} of a parent patch and each of its eight child patches, this numerical error can be made diminished in Eq. (\ref{eq:CorrectedPot}).

To further improve the solution across the inter-patch boundaries, a \emph{sibling relaxation step} is employed immediately after solving Eqs. (\ref{eq:Poisson}) and (\ref{eq:PotCorrection}). In this step, instead of using the interpolated values as the fine-grid boundary conditions, we exchange the values near the inter-patch boundaries between neighboring patches, store them in the corresponding ghost zones, and apply another SOR iteration. Finally, when necessary, a few times of the sibling relaxation steps may also be employed after Eq. (\ref{eq:CorrectedPot}). Nevertheless, it only gives rise to minor corrections to the final results. In \emph{GAMER}, typically we apply one sibling relaxation step after solving Eqs. (\ref{eq:Poisson}) and (\ref{eq:PotCorrection}), and $2\sim 4$ sibling relaxation steps after Eq. (\ref{eq:CorrectedPot}).

The procedure of the two-level solution described above can be straightforwardly generalized to the multi-level case. It works in the steps as follows.
\begin{enumerate}
\item In each level $\ell$ from the root level ($\ell=0$) up to the maximum level $\ell_{max}$, evaluate the uncorrected potential $\phi_{uncorrected}$ by solving Eq. (\ref{eq:Poisson}). Immediately after that, perform one sibling relaxation step.
\item In each level $\ell$ from level $\ell_{max}-1$ down to level $0$, evaluate the pseudo mass sheet $\xi$ via Eqs. $(\ref{eq:PseudoRho})-(\ref{eq:FineGradientPhi})$, and add up with the pseudo mass sheet derived from patches at level $\ell+1$
\item In each level $\ell$ from level $0$ up to level $\ell_{max}$, evaluate the correction $\zeta$ by solving Eq. (\ref{eq:PotCorrection}). Immediately after that, perform one sibling relaxation step.
\item For each level $\ell$ from level $0$ up to level $\ell_{max}$, obtain the corrected potential $\phi_{corrected}$ by Eq. (\ref{eq:CorrectedPot}). Afterwards, if necessary, perform a few times of sibling relaxation steps.
\end{enumerate}

Note that this multi-level procedure is similar to the scheme proposed by \citet{Ricker08}. However, instead of using the residuals as the source function to solve the potential correction (as applied in the Ricker's scheme), our method evaluates the potential correction via estimating the pseudo mass sheet and features the conservation of mass. This feature is described below in more detail.

The potential correction scheme adopted in \emph{GAMER} has two importance nice features. First, since the Poisson equation is solved in individual patches with inhomogeneous Dirichlet boundary conditions, the inter-patch communication is minimized, and all patches at a given level can be solved independently. By taking advantage of this feature, we can implement a high-performance GPU Poisson solver, in which hundreds of patches can be solved in parallel. The second and most critical feature is the conservation of mass. Since the pseudo mass sheet is defined by the jump of the normal derivative of potential, our scheme enforces the summation of local pseudo mass sheet introduced by the interface between a parent patch and a child patch to be zero (to the machine precision).

The feature of mass conservation can be easily seen from the fact that both the coarse-grid and fine-grid potentials satisfy the Poisson equation. Therefore, the Gauss's theorem states that
\begin{equation}
\oint_\Gamma\left(\left.\frac{\partial\phi}{\partial n}\right|_{\epsilon^-}\right)ds=4\pi GM^c ~~~ {\rm and} ~~~
\oint_\Gamma\left(\left.\frac{\partial\phi}{\partial n}\right|_{\epsilon^+}\right)ds=4\pi GM^f,
\label{eq:Gauss}
\end{equation}
where $\Gamma$ is the closed boundary surface of a child patch, and $M^c$ and $M^f$ are the total coarse-grid mass and fine-grid mass enclosed by $\Gamma$, respectively. Now, in \emph{GAMER} we always have $M^c=M^f$ thanks to the constrained operation, and hence
\begin{equation}
\oint_\Gamma\xi({\bf x})ds=\frac{1}{4\pi G\triangle h^c}\oint_\Gamma\left(\left.\frac{\partial\phi}
{\partial n}\right|_{\epsilon^-}-\left.\frac{\partial\phi}{\partial n}\right|_{\epsilon^+}\right)ds=0.
\label{eq:ZeroPseudoRho}
\end{equation}
Clearly, Eqs. (\ref{eq:Gauss}) and (\ref{eq:ZeroPseudoRho}) still hold in their discrete forms, in which the normal derivatives of potentials are approximated by Eqs. (\ref{eq:CoarseGradientPhi}) and (\ref{eq:FineGradientPhi}).

The physical picture for reinforcing the summation of local pseudo mass sheet to vanish is as follows. Beside the truncation error in Eq. (\ref{eq:DiscretePoisson}), the numerical error in the fine-grid region is caused by the insufficiently accurate boundary condition, which is obtained by interpolation on the coarse-grid values. In other words, even though the refined region can provide higher resolution, it does not contribute to the coarse-grid solution. The absence of local density distribution in the fine-grid region when evaluating the coarse-grid potential can only provide the coarse-grid resolution, and the numerical error will propagate into the solution of fine-grid potential through the setting of fine-grid boundary condition (even with high-order interpolation). It is the reason that we want to estimate the pseudo mass sheet to correct the coarse-grid potential, and thereby provide a more accurate boundary condition for solving the fine-grid potential. However, since in \emph{GAMER} the total mass within a given volume is the same in different refinement levels, there should be \emph{no mass monopole correction} for the coarse-grid potential. Therefore, the total pseudo mass introduced by the coarse-fine interface between a parent patch and a child patch is zero.

Finally, the cell-centered gravitational accelerations are evaluated by the 3-points finite difference approximation of the gradient operator. The flow variables are advanced by solving Eqs. (\ref{eq:MomConserve}) and (\ref{eq:EnergyConserve}), in which the flux terms are ignored. By utilizing the same operator-splitting method described in \S\ 2.2, the Euler equations with self-gravity can be solved in the order $xyzGGzyx$ \citep{TP03}, in which the operator $xyz$ represents the order of directions to update flow variables by the flux differences, and the operator $G$ represents the updating of flow variables by gravity. Note that the continuity equation (Eq. [\ref{eq:MassConserve}]) has no self-gravity term. Therefore, the two successive self-gravity operators $GG$ can be combined together with a twice larger time step.

In cosmological simulations, the Poisson equation can be rewritten as
\begin{equation}
\tilde\nabla^2\tilde\phi({\bf \tilde x})=4\pi Ga[\tilde\rho({\bf \tilde x})-\tilde\rho_b({\bf \tilde x})],
\label{eq:ComovingPoisson}
\end{equation}
where each comoving variable is defined in Eqs. (\ref{eq:ChangeOfVariable1}) and (\ref{eq:ChangeOfVariable2}), and the subscript $b$ indicates the background density. The gravitational constant $G$ can be replaced by
\begin{equation}
G=\frac{3H_0^2\Omega_{m,0}}{8\pi\rho_{b,0}},
\label{eq:ComovingG}
\end{equation}
where $\Omega_{m,0}$ is the matter density and the subscript $0$ indicates the values at the present time.

\subsection{Parallelization}

For astrophysical simulations, the spatial resolutions are generally limited by the amount of total memory. It is therefore essential to develop a parallel code that distributes the workload to multiple processors. Accordingly, \emph{GAMER} is developed to work in a multi-CPU system with multi-GPU acceleration. Each CPU manages one GPU, and the data transfer between different CPUs is accomplished by using the MPI library.

Developing a parallel GPU-accelerated program requires elaborate treatments. First of all, even though the computation time may be highly reduced by using GPUs, \emph{the communication time is not reduced at all}. The network bandwidth may easily become the performance bottleneck, and results in a limited overall performance improvement. Moreover, since the performance of the GPU solver also relies on the massively parallel architecture inside GPU, our parallel algorithm must preserve this capability.

\begin{figure}
\centering
\includegraphics[width=15cm]{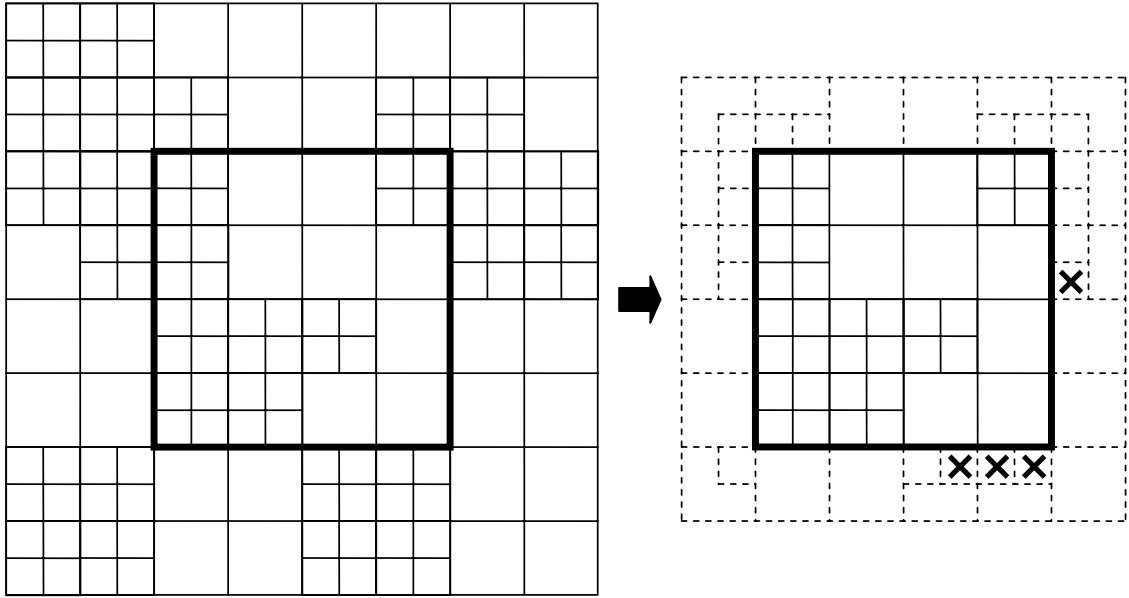}
\caption{Two-dimensional example of the allocation of buffer patches with two refinement levels. The left figure shows the refinement map within a fixed domain. Each grid represents a patch, and the bold contour represents the boundaries of a sub-domain. The right figure shows the patches actually allocated to the processor in charge of the sub-domain. The solid and dashed lines represent the real patches and buffer patches, respectively. The crosses indicate the patches that do not require to store physical quantities.}
\label{fig:BufferPatch}
\end{figure}

In \emph{GAMER}, the parallelism is based on a rectangular domain decomposition. All patches within a rectangular sub-domain are calculated by one CPU/GPU combination. These patches are referred as the \emph{real patches}. Boundary conditions of each sub-domain are provided by allocating the \emph{buffer patches} surrounding the sub-domain. Figure \ref{fig:BufferPatch} shows a two-dimensional example of the allocation of buffer patches. The physical data stored in each buffer patch are always filled by transferring data between processors. Note that each buffer patch also stores the correct indices of the parent, child, and 26 sibling patches, and a null value is assigned if the corresponding patch does not exist. Accordingly, all coarse-fine boundaries can be correctly identified even if they coincide with the sub-domain boundaries. Moreover, by doing so, we do not need to store all the oct-tree data structure redundantly in all processors.

To avoid the bottleneck of network communication, the amount of data transfer between different CPUs is carefully minimized in \emph{GAMER}. For the hydrodynamic solver, since the TVD scheme requires three ghost-zone values, only a three-cells-wide array is transferred and stored in each buffer patch. The data stored in the buffer patches at level $\ell$ are used to set the ghost-zone values for both the patches at level $\ell$ via direct memory copy as well as the patches at level $\ell+1$ via interpolation. If a coarse-fine interface coincides with the sub-domain boundaries, the corresponding flux data are also transferred for the flux correction operation. For the Poisson solver, since it requires only one ghost-zone value, a two-cells-wide array is transferred for setting the boundary conditions via interpolation, and an one-cell-wide array is transferred for the sibling relaxation step.

We note that not all buffer patches are necessary to be filled with the hydrodynamic and potential data. To be more precise, a buffer patch is required to receive data only if any of its 26 sibling patches corresponds to a real patch. A two-dimensional example is also illustrated in Figure \ref{fig:BufferPatch}, in which the buffer patches marked with a cross do not required to store physical quantities and are used only to provide the correct oct-tree data structure for real patches adjacent to the sub-domain boundaries. Also note that the additional memory overhead associated with the allocation of buffer patches is generally negligible as long as the number of real patches is sufficiently large.

Adopting the FFTW library as the root-level Poisson solver requires additional works. The parallelization strategy of FFTW is based on the slab decomposition, which is incompatible with the rectangle decomposition adopted by \emph{GAMER}. Therefore, a rectangle-to-slab transformation must be performed before the root-level Poisson solver, and a slab-to-rectangle transformation must be performed after the root-level Poisson solver. Both these two operations require global communications. However, since in general the amount of data in the root level is much less than that in higher levels, this additional communication time is usually negligible.

The parallelization algorithm described above requires a recurrent search for all patches within the sub-domain. It can also be time-consuming in a large-scale simulation, especially when the performance of single-patch solvers are highly improved by using GPUs. In \emph{GAMER}, we get around this problem by first constructing a table recording the indices of root-level \emph{border patches}, which is defined as the real patches adjacent to each side of the sub-domain boundaries. The table recording the indices of ``higher-level'' border patches can then be constructed hierarchically, as a consequence of the fact that a border patch at level ${\ell}$ must be the child patch of a border patch at level $\ell-1$. After that, the table listing the indices of patches to send and to receive data can be built by only searching over the border-patch table. Therefore, the global search is performed only at the root level, at which the number of patches is much smaller than that of higher levels. Also note that these tables are only re-constructed every time after rebuilding the refinement map; afterwards they can be reused until the next domain refinement.

As the number of processors increases, applying the rectangular domain decomposition in the AMR implementation can lead to an issue of load unbalance, where different computing nodes have different loads. Generally, this issue is solved by adopting the method of space-filling curve \citep[e.g.,][]{Campbell03} to redistribute the computing loads. This is currently being implemented into the \emph{GAMER} code. Note that the concept of allocating the buffer patches should still be adopted, as long as we impose the constraint that each child patch is placed in the same node (more precisely the same MPI rank) as its parent patch. Preliminary tests have shown that this constraint results in only a minor influence of the load balance. The unbalance is found to be less than 2\% in a purely-baryonic cosmological simulation with 32 CPU/GPUs.

\section{GPU IMPLEMENTATION}

In \emph{GAMER}, the GPU implementation is inspired by the two parallelism levels naturally embedded inside the AMR structure. First, each mesh patch can be calculated independently as long as its ghost-zone values are provided. Therefore, we can use one thread block to calculate one patch. Second, all cells inside a patch can be calculated in parallel as long as there is a synchronization mechanism to coordinate the data update of each cell. This can be accomplished by using multiple CUDA threads to calculate different cells within the same patch, and store the updated results in the shared memory. In the following, we describe in detail the GPU implementations of different parts in the code, and give the results of timing analyses.

\subsection{GPU hydrodynamic solver}

The GPU hydrodynamic solver implemented in \emph{GAMER} involves three basic steps as follows.
\begin{enumerate}
\item Send the input array, which stores mass density, momentum density, and energy density, downstream from the CPU memory to the GPU global memory.
\item Invoke the GPU hydrodynamic kernel to advance the equations for one time step.
\item Send the output array, which stores mass density, momentum density, and energy density, upstream from the GPU global memory back to the CPU memory.
\end{enumerate}

Before invoking the GPU kernel, a \emph{preparation step} is performed by CPU to prepare the input array which stores the interior data of each patch and the ghost-zone values. For the TVD scheme, three ghost zones are required on each side in each spatial direction. The ghost-zone values are obtained by direct memory copies if the sibling patches exist. Otherwise, the values are obtained by linear interpolation with the Min-Mod limiter.

Note that eight nearby patches are always allocated simultaneously thanks to the oct-tree data structure. Therefore, to reduce the amount of workload associated with the preparation of the ghost-zone values, we can group these eight patches into a larger array (hereafter referred as the \emph{patch group}) before sending into GPU. For the hydrodynamic solver, each patch group contains $(8\times2+3\times2)^3=22^3$ cells, where 8 is the size of a single patch and 3 is the size of the ghost zones. Since in this approach the ghost zones are prepared for the exterior region of each patch group instead of each individual patch, it reduces 63 percent of the computational overhead associated with the ghost-zone preparation.

After sending the input array into GPU, each patch group is advanced by one GPU thread block. The single-block GPU hydrodynamic solver is based on the TVD scheme described in \S\ 2.2. The three-dimensional evolution is achieved by using the dimensional splitting method, in which the solution is obtained by first applying a forward sweep followed by a backward sweep within the same step. During one GPU kernel execution, either a forward sweep or a backward sweep is performed, and a boolean parameter is sent into GPU to indicate the direction of sweeping.

Since the dimensional-split Euler equations are equivalent to a set of one-dimensional conservation equations, the data of a single patch group can be regarded as a set of data columns, and each of which can be evolved independently. Accordingly, for the single-block GPU hydrodynamic solver, we advance the solutions of a fixed number of data columns in parallel. Each thread is responsible for advancing a single cell. We then iterate through the remaining data columns until the whole patch group is updated. Data that need to be accessed by more than one threads within the same data column (e.g., the fluid fluxes) are stored in the GPU shared memory. Otherwise they are stored in the per-thread registers.

Briefly, the GPU hydrodynamic kernel executed by each thread works with the steps as follows.
\begin{enumerate}
\item Get the index of cell being calculated. Fetch the corresponding data from the global memory and store in the per-thread registers.
\item Calculate the freezing speed $c(x,t)$ and construct the corresponding auxiliary vector ${\bf w}$.
\item Calculate the left-moving and right-moving fluxes (Eq. [\ref{eq:LRFlux}]) defined at the boundaries of each cell by the first-order upwind scheme. Store the fluxes in the shared memory.
\item Obtain the midpoint solutions by Eq. (\ref{eq:MidpointValue}). Store the solutions in the per-thread registers.
\item Recalculate the freezing speed using the midpoint values. Construct the corresponding midpoint auxiliary vector.
\item Calculate the midpoint left-moving and right-moving fluxes defined at the boundaries of each cell by the second-order TVD scheme. Store the fluxes in the shared memory.
\item Obtain the full-step solutions by Eq. (\ref{eq:FullstepValue}). Store the solutions in the per-thread registers.
       \item Store the full-step solutions as well as the fluxes across the patch boundaries back to the global memory.
\item Repeat steps $1-8$ for the next targeted data column until the entire patch group is updated.
\item Repeat steps $1-9$ for the next one-dimensional sweep until either a forward sweep or a backward sweep is complete.
\end{enumerate}

The output array stores the updated solutions of each patch group as well as the fluxes across the boundaries of each patch. To reduce the amount of data transfer, no ghost-zone values are stored in the output array. After the GPU kernel execution, the output array is transferred upstream to the CPU memory, and followed by a \emph{closing step} performed by CPU.

The closing step involves two operations. First, it copies the updated solutions back to each corresponding patch pointer. Second, for a patch adjacent to a coarse-fine interface, the data of fluxes across this interface are copied into its own flux array (to be corrected afterwards by the fine-grid fluxes) if this patch is in the coarse side of the interface. On the contrary, if this patch is in the fine side of the interface, the data of fluxes are copied into the flux array of the corresponding neighboring coarse patch (to correct the coarse-grid fluxes).

We notice that it is unnecessary to simultaneously advance all patch groups in GPU, since the GPU computing power can be fully exploited as long as there are sufficient arithmetic operations. Typically, we advance $128-240$ patch groups in GPU in parallel. The input and output arrays are allocated only for the patch groups being updated, and a single array can be reused for different sets of patch groups. The additional memory requirement in CPU for storing the ghost-zone data is therefore nearly negligible. Moreover, the memory requirement in GPU for the hydrodynamic kernel is less than 200 MB, and hence the limited amount of DRAM memory in GPU is not an issue in the current implementation of \emph{GAMER}.

In a practical point of view, the performance comparisons between CPU and GPU must include the data transfer time between the CPU memory and the GPU memory through the PCI Express bus. This data transfer time can be greatly reduced by utilizing the capability of asynchronous memory copies in GPU, in which the memory copies between CPU and GPU can be overlapped with the kernel executions \citep{NVIDIA08}. In CUDA, the concurrency between different operations is managed by creating the \emph{stream objects}, which contain a sequence of memory copy operations and kernel invocations. To simplify the discussion, here we assume that a stream identification number (hereafter referred as the \emph{stream ID}) is assigned to each stream object. For a memory copy operation and a kernel launch with different stream IDs, they can be performed concurrently. Otherwise, they will be performed sequentially in the order they are declared in the same stream object.

In \emph{GAMER}, since different patch groups at the same refinement level can be evolved in an arbitrary order, we can create several stream objects inside one GPU solver. Each stream object contains one downstream memory copy, one kernel invocation, and one upstream memory copy for a fixed number of patch groups. Accordingly, the data transfer time of the patch groups belonging to one stream object can be overlapped with the kernel execution time of the patch groups belonging to a different stream object.

\begin{figure}
\centering
\includegraphics[width=15cm]{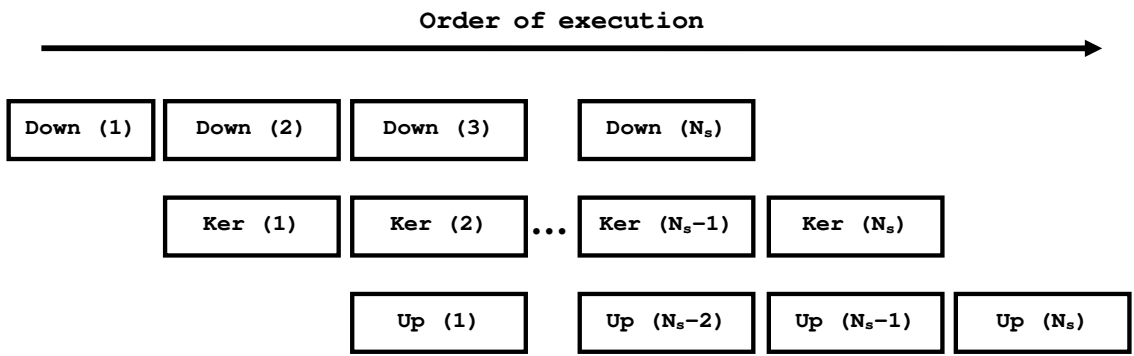}
\caption{Illustration of the concurrent execution between the memory copy and the kernel launch. The abbreviations ``Down'', ``Up'', and ``Ker'' stand for the downstream memory copy, upstream memory copy, and kernel execution, respectively. The numbers in the parentheses indicate the stream IDs. The operations at the same column are performed concurrently. Therefore, the memory copies can be overlapped with the kernel executions.}
\label{fig:AsyncMemcpy}
\end{figure}

As an illustration, let $N_s$ denote the number of stream objects and $N_g$ denote the number of patch groups associated with one stream object. The total number of patch groups advanced by one launch of the GPU solver is then given by $N_s\times N_g$. At the first step, the patch groups 1 to $N_g$ are sent to the GPU memory. At the second step, a GPU kernel is executed to advance the patch groups 1 to $N_g$, while at the same time the patch groups $N_g\!+\!1$ to $2N_g$ are sent to the GPU memory. At the third step, a GPU kernel is executed to advance the patch groups $N_g\!+\!1$ to $2N_g$, while at the same time the updated solutions of the patch groups 1 to $N_g$ are sent back upstream to the CPU memory, and the prepared data of the patch groups $2N_g\!+\!1$ to $3N_g$ are sent downstream to the GPU memory. Figure \ref{fig:AsyncMemcpy} illustrates the complete procedure. $N_s\!+\!2$ steps are required to complete one launch of the GPU solver. Theoretically, by using $N_s$ streams, the data transfer time between CPU and GPU can be reduced to ($1/N_s$)-th of the time without using streams.

For comparison, a CPU hydrodynamic solver with the same TVD scheme has also been implemented. In order to have more reliable timing measurements, we have measured the performance of the hydrodynamic solver in three different hardware implementations: the GraCCA system, the GPU system installed in the National Center for High-Performance Computing of Taiwan (hereafter referred as NCHC), and the GPU system installed in the Center for Quantum Science and Engineering of National Taiwan University (hereafter referred as CQSE). Below, we give a brief description of the hardware implementations in different GPU systems.

The GraCCA system contains 18 nodes connected by Gigabit Ethernet. Each node is equipped with two GeForce 8800 GTX GPUs and one AMD Athlon 64 X2 3800 CPU. Two distinct configurations of GPU systems are implemented in NCHC. First, a GPU cluster consisting of 16 nodes connected by InfiniBand is installed in NCHC. In order to exploit the bandwidth of InfiniBand, currently each node is only equipped with two Tesla T10 GPUs and two Intel Xeon X5472 CPUs. In addition, an experimental node with four Tesla T10 GPUs and two Intel Xeon E5520 CPUs is also installed in NCHC. This node aims at exploring the computing power of four Tesla GPUs. Therefore, throughout this work, we perform the timing measurements in NCHC at this node. The CQSE GPU cluster contains 16 nodes connected by Gigabit Ethernet. Each node is equipped with four Tesla T10 GPUs and two Intel Xeon E5462 CPUs.

Note that in contrast to the Tesla T10 GPU, the GeForce 8800 GTX GPU installed in the GraCCA system does not support the capability of asynchronous memory copies. Consequently, the memory copies and the kernel invocations must be performed sequentially, which is equivalent to have only one stream. For the Tesla T10 GPU, we have compared the performances between tests using only one stream ($N_s=1$) and four streams ($N_s=4$). Also note that each thread block is calculated by one multiprocessor in GPU, and there are 16 and 30 multiprocessors in GeForce 8800 GTX GPU and Tesla T10 GPU, respectively.

\begin{figure}
\centering
\includegraphics[width=15cm]{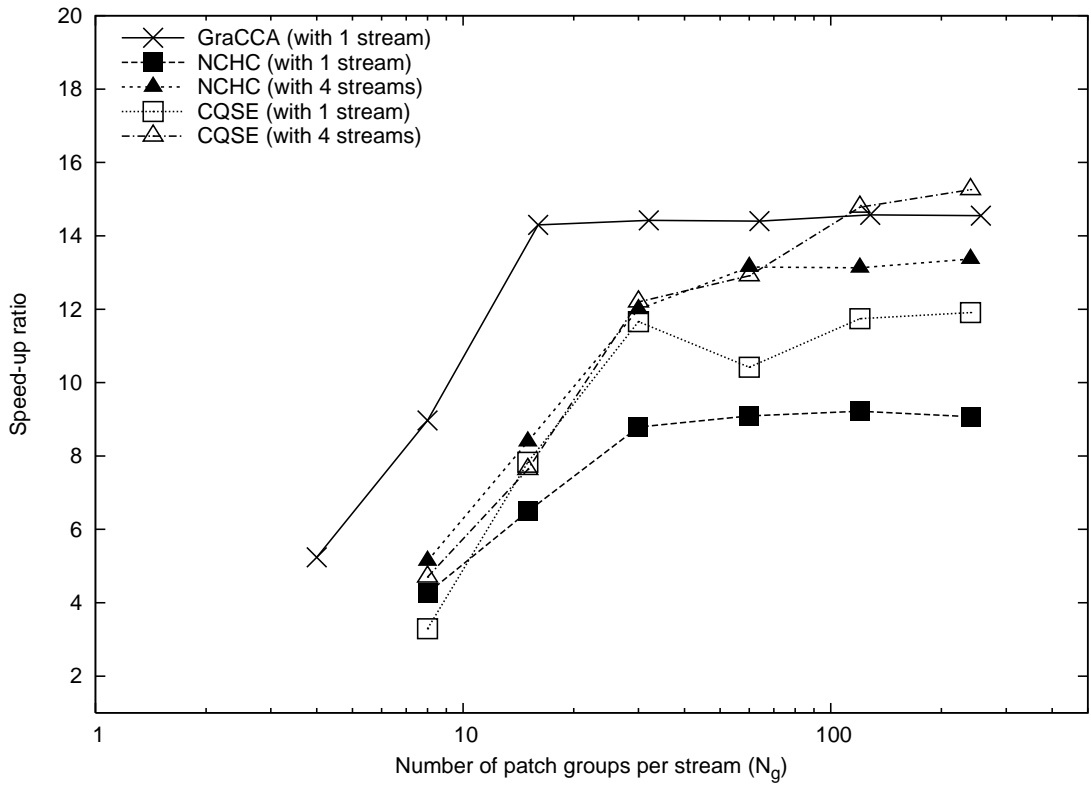}
\caption{Performances of the GPU hydrodynamic solver in different hardware implementations.}
\label{fig:GPUFluidSolver_GPUvsCPU}
\end{figure}

Figure \ref{fig:GPUFluidSolver_GPUvsCPU} shows the performance speed-up of one GPU over one CPU as a function of the number of patch groups associated with each stream ($N_g$) for the hydrodynamic solver. The performance measurements include the downstream and upstream data transfers as well as the kernel executions. The computing times for the preparation step and the closing step are not included. It can be seen that the performance of GPU solver is linear proportional to $N_g$ when $N_g$ is smaller than the total number of multiprocessors in one GPU. As we have at least one patch group for each multiprocessor, the performance approaches the saturated values. Factors of 14.6, 13.4, and 15.3 performance speed-ups are demonstrated in the GraCCA system, NCHC, and CQSE, respectively, for GPU computation as opposed to CPU computation.

A detailed timing analysis for the hydrodynamic solver is listed in Table \ref{table:TimingHydro}, in which we set $N_g=256$ in the GraCCA system and $N_g=240$ in NCHC and CQSE. Note that since the specifications of GPUs installed in NCHC and CQSE are the same, the difference of speed-up ratios between these two systems is mainly due to the different performances of CPUs as well as the different Northbridge chips. When using only one stream, the data transfer times are 46.4\%, 79.9\%, and 48.7\% of the kernel execution times in the GraCCA system, NCHC, and CQSE, respectively. A relatively low bandwidth in PCI Express bus is found in NCHC, especially in the upstream bandwidth. Nevertheless, the data transfer times are reduced to 22.9\% in NCHC and 18.0\% in CQSE when the memory copies are partially overlapped with the kernel executions by using four streams. Also note that both GPUs and CPUs installed in NCHC and CQSE outperform those installed in the GraCCA system. However, the performance ratios between GPU and CPU measured in different hardware implementations do not vary significantly.

\subsection{GPU Poisson Solver}

The basic procedure of the GPU Poisson solver is similar to that of the GPU hydrodynamic solver. It works with the steps as follows.
\begin{enumerate}
\item Send the input array, which stores mass density and potential, downstream from the CPU memory to the GPU global memory.
\item Invoke the GPU SOR kernel to evaluate the potential solutions.
\item Send the output array, which stores potential only, upstream from the GPU global memory back to the CPU memory.
\end{enumerate}

In order to reduce the computational overhead associated with the ghost-zone preparation as well as to reduce the interfaces between patches at the same refinement level, eight nearby patches are also grouped into a patch group at the preparation step. For the SOR method, the potential data require one ghost zone on each side in each spatial direction, while the mass density data require no ghost zones. Accordingly, for the GPU Poisson solver, each patch group stores $(8\times2+1\times2)^3=18^3$ potential data and $(8\times2)^3=16^3$ density data. After the preparation step, the input array is sent downstream to the GPU global memory, and the GPU SOR kernel is invoked to evaluate the potential solutions of each patch group. Afterwards, the output array storing only the potential solutions are sent upstream to the CPU memory, and a closing step is performed by CPU to copy the solutions back to each corresponding patch pointer.

The single-block GPU Poisson solver is based on the SOR scheme with odd-even ordering as described in \S\ 2.3. Implementing the three-dimensional SOR scheme into GPU differs greatly from the implementation of the three-dimensional TVD scheme. For the GPU hydrodynamic kernel, the data of each patch group are decomposed into a set of data columns, and each of which can be evolved independently. Accordingly, only the data columns being calculated need to be stored in the shared memory, and a single shared memory array can be reused for many different data columns. The total amount of shared memory required in the GPU hydrodynamic kernel is therefore only 8.6 KB.

On the contrary, the SOR scheme requires the three-dimensional relaxation, where the data in each cell must be re-accessed during each iteration. In addition, the arithmetic intensity in each iteration of the SOR scheme is relatively low. Consequently, storing all data naively in the global memory will suffer from the high memory latency and result in marginal performance improvement. However, since the amount of potential data in each patch group is about 22.8 KB, it is impossible to store all potential data in the shared memory (which is only 16 KB per multiprocessor).

One of the solutions to reduce the data transfer is that during each iteration, we solve Eqs. (\ref{eq:Residual}) and (\ref{eq:SOR}) slice-by-slice. Since Eq. (\ref{eq:Residual}) requires the data of six nearby cells, we can keep only three slices of data in the shared memory, and fetch a new slice into the shared memory each time when iterating to the next slice. It ensures that the reused data are stored in the shared memory during each iteration. However, this method still requires frequent data transfer between the global memory and the shard memory, and hence the achieved performance is far from optimized. On the contrary, in \emph{GAMER} we have implemented a different scheme that minimizes the data transfer by utilizing both the shared memory and the per-thread registers, as detailed below.

To solve the issue of shared memory shortage, we first notice that there are 8,192 and 16,384 threads per multiprocessor in the GeForce 8800 GTX and Tesla T10 GPU, respectively. In principle, they provide another 32 KB and 64 KB storage through the per-thread registers, although the data stored in one register cannot be directly accessed by other registers. Next, by dividing all cells within a single patch group into odd and even cells, the update of an even cell depends only on nearby odd cells, and vice versa. Accordingly, when updating all even cells at the even step in one iteration, \emph{the data of all even cells can be stored in the per-thread registers} instead of the shard memory. By contrast, since the data of each odd cell need to accessed several times by nearby even cells, \emph{the data of all odd cells are stored in the shared memory} at the even step. After updating all even cells, we exchange the data stored in the registers and the shared memory, and proceed to the odd step for updating all odd cells. At the end of the odd step, another data exchange operation is applied and a single iteration is complete.

\begin{figure}
\centering
\includegraphics[width=8cm]{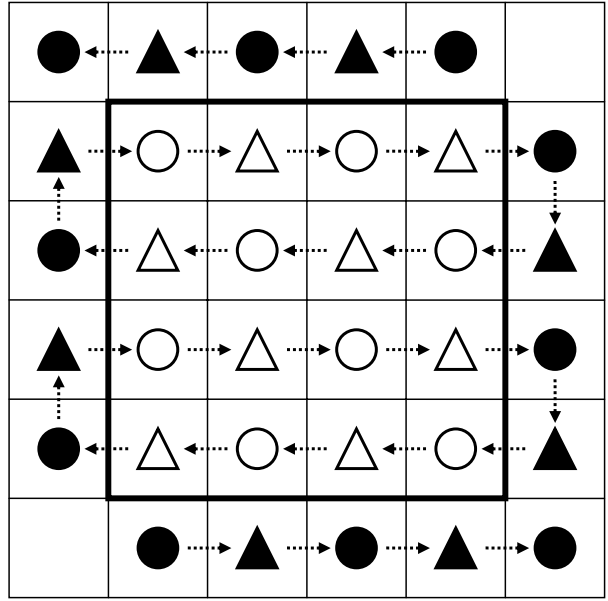}
\caption{Two-dimensional illustration of the configuration of odd and even cells in the GPU SOR kernel. To simply the illustration, a patch group with only $4\times4$ cells is shown. The border of patch group is highlighted by bold lines. The triangles and circles represent the even and odd cells, respectively. The interior cells are indicated by open symbols, and the ghost cells are indicated by filled symbols. At the first half-step for updating all even cells, the triangle data are stored in the per-thread registers, while the circle data are stored in the shared memory. A shared memory array with $6\times3$ elements is allocated, and 8 threads are used to solve the equation. Each thread stores the potentials of one interior cell and one ghost cell in its own registers. The dotted arrows indicate the direction of data exchange at the end of the even step.}
\label{fig:GPU_SOR}
\end{figure}

Figure \ref{fig:GPU_SOR} illustrates the configuration of odd and even cells as well as the data exchange operation at the even step in two dimensions. Note that the ghost zones are also divided into odd and even cells, except that their values are fixed during iterations. In this scheme, the requirement for the amount of shared memory is nearly halved, and all potential data are stored in either the shared memory or the per-thread registers. The data transfers of the potential data between the global memory and the shared memory are only necessary before and after the relaxation loop, and thus the performance is highly improved. Also note that since the access rate of the mass density is much lower than that of the potential, the data of mass density are still stored in the global memory.

According to the discussions above, we store the potential data of each patch group in either the shared memory or the per-thread registers, and update the potential solutions slice-by-slice. Since there are 128 interior even cells and 128 interior odd cells in each x-y plane in one patch group, we use 128 threads for each patch group. Each thread will update the potential of one even cell in each slice (and store in the register) at the even step and update the potential of one odd cell in each slice (and store in the register) at the odd step. Since there are 16 slices of potential data to be updated in each patch group, each thread requires 16 registers for storing the temporary interior potentials. Besides, each thread also requires 6 registers for storing the ghost-zone potentials on each side of the patch group. The GPU SOR kernel executed by each thread works with the steps as follows.
\begin{enumerate}
\item Load the potentials of odd cells into the shared memory.
\item Load the potentials of even cells into the per-thread registers.
\item Initialize the parameters for the even step.
\item Evaluate the residual of an even cell at the targeted x-y plane by Eq. (\ref{eq:Residual}). Store the residual in a shared memory array.
\item Update the solution of an even cell by Eq. (\ref{eq:SOR}).
\item Repeat steps $4-5$ for the next x-y plane until all even cells are updated.
\item Exchange the data stored in the shared memory and the per-thread registers.
\item Initialize the parameters for the odd step. Repeat steps $4-7$ for all odd cells.
\item Perform a reduction operation to get the 1-norm of the residuals. Repeat steps $3-9$ until the 1-norm of the residuals has been diminished to the floating-point precision limit (compared to the source term in the Poisson equation).
\item Store the solutions back to the global memory.
\end{enumerate}

\begin{figure}
\centering
\includegraphics[width=15cm]{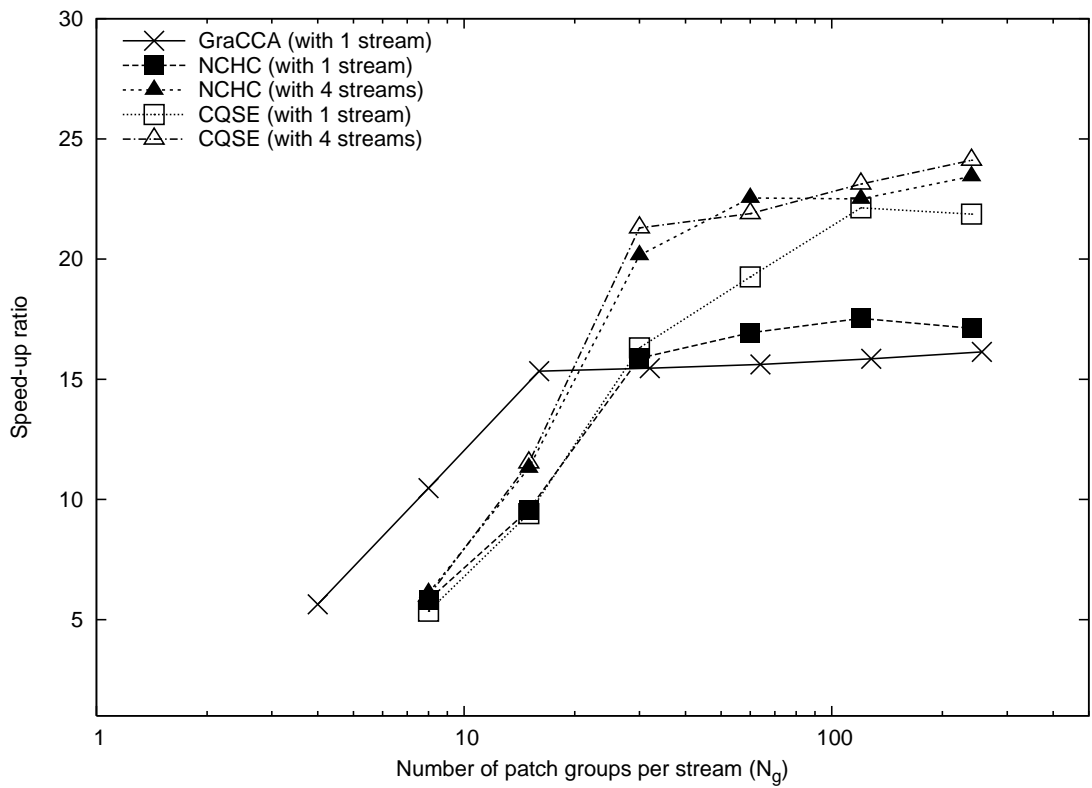}
\caption{Performances of the GPU Poisson solver in different hardware implementations.}
\label{fig:GPUPoissonSolver_GPUvsCPU}
\end{figure}

A CPU Poisson solver with the SOR scheme has been implemented for the sake of comparison. Note that the data exchange operation is not required in the CPU solver. Figure \ref{fig:GPUPoissonSolver_GPUvsCPU} shows the performance speed-up of one GPU over one CPU as a function of the number of patch groups associated with each stream ($N_g$) for the Poisson solver. The performance measurements include the downstream and upstream data transfers as well as the kernel executions, whereas the computing times for the preparation step and the closing step are not included. In the NCHC and CQSE systems, four streams are used for the asynchronous memory copies. For comparison, the results using only one stream are shown as well. As for the GPU hydrodynamic solver, the speed-up ratios are linear proportional to $N_g$ when $N_g$ is smaller than the number of multiprocessors in one GPU, and they approach the saturated values when we have at least one patch group for each multiprocessor. Factors of 16.1, 23.4, and 24.1 performance speed-ups are demonstrated in the GraCCA system, NCHC, and CQSE, respectively.

Table \ref{table:TimingPoisson} shows a detailed timing analysis for the Poisson solver, in which we set $N_g=256$ in the GraCCA system and $N_g=240$ in NCHC and CQSE. Again, a relatively low bandwidth of the PCI Express bus is measured in NCHC. When using only one stream, the data transfer times are 20.0\%, 56.4\%, and 33.3\% of the kernel execution times in the GraCCA system, NCHC, and CQSE, respectively, and they are reduced to 15.4\% in NCHC and 21.4\% in CQSE as the asynchronous memory copies are activated by using four streams.

\subsection{Gravitational Acceleration}

In \emph{GAMER}, the evaluation of the potential gradients as well as the updating of hydrodynamic variables by self gravity are also performed in GPU. The procedure is analogous to the ones adopted in the hydrodynamic and Poisson solvers. An input array storing both the hydrodynamic variables and the potentials is sent downstream to GPU, a GPU kernel is executed to update the solutions, and the updated hydrodynamic variables are sent upstream back to CPU.

Note that the arithmetic intensity is extremely low in this case, and hence the data transfer time between CPU and GPU dominates the execution time. Consequently, only marginal performance speed-up factors around $1.3\sim 1.4$ are achieved in the GraCCA system. In NCHC and CQSE, the performances are even slightly lower than using CPU. Nevertheless, in the cosmological tests using CPU only, the execution time associated with the calculation of gravitational acceleration only accounts for $\sim$ 1\% of the total execution time. Therefore, the overall performance of the code is not constrained by the relatively poor performance of the computation of gravitational acceleration.

\subsection{Courant-Friedrichs-Lewy Condition}

To ensure the stability of numerical integration, the integration time step must satisfy the Courant-Friedrichs-Lewy (CFL) stability criterion. For the three-dimensional relaxing TVD scheme, the CFL condition is given by
\begin{equation}
\triangle t\leq \frac{\triangle h_{\ell}}{c_s+max(|v_x|,|v_y|,|v_z|)},
\label{eq:CFL}
\end{equation}
where $c_s$ is the sound speed and $\triangle h_{\ell}$ is the zone spacing at level $\ell$. The denominator in Eq. (\ref{eq:CFL}) gives the maximum speed of information propagation.

The evaluation of the CFL condition may, at first sight, seem to take negligible time compared to the total execution time. However, timing measurements show that, after the performances of both the hydrodynamic and Poisson solvers are highly improved by using GPUs, the calculation time for the CFL condition accounts for at most 8\% of the total execution time if the CFL condition is computed in CPU. It is comparatively low but non-negligible.

Due to the fact that the evaluation of the CFL condition requires no ghost zones, we can calculate the CFL condition in GPU after the backward sweep of the GPU hydrodynamic solver. Since all data essential for the CFL condition already reside in the GPU memory, no extra downstream data transfer is required. To reduce the amount of upstream data transfer for the CFL condition, a reduction operation is performed in GPU in advance to obtain the maximum speed of information propagation among each patch group. By doing so, only a single floating-point value per patch group is required to be transferred from GPU to CPU. The reduction operation among different patch groups are then performed by CPU to evaluate the CFL condition. Accordingly, since the extra data transfer is minimized and the number of arithmetic operations of the CFL condition is much less than that of the hydrodynamic solver, the computation time of the CFL condition is reduced to less than 0.1\% of the total.

\subsection{Concurrent Execution between CPU and GPU}

In \emph{GAMER}, two optimization strategies are adopted to minimize the computational overhead introduced by the preparation step and the closing step. First, as described in \S\ 3.1 and \S\ 3.2, we prepare the ghost-zone data for each patch group instead of each patch. The surface/volume ratio is reduced and thus the computational overhead is reduced. However, since the number of the interior cells in a single patch group is only $16^3$, the number of the ghost zones is still comparable to that of the interior cells. For example, for the hydrodynamic solver which requires three ghost zones, the total number of ghost zones for each patch group is 1.6 times more than the number of interior cells. Moreover, although the number of arithmetic operations involved in the preparation step and the closing step are much less than that of the hydrodynamic and Poisson solvers, \emph{these two steps are performed by CPU} in the current scheme. On the contrary, the execution times of both solvers are reduced by at least a factor of ten by using GPU. Consequently, timing analyses show that the execution times of the preparation step and the closing step are comparable to or even longer than that of the GPU solvers.

To further remove this bottleneck, a second optimization strategy is implemented in \emph{GAMER} by taking advantage of the parallel execution between CPU and GPU. In CUDA, both the memory copy operations and the kernel invocations in GPU are asynchronous, meaning that the program returns from the function call before the requested task is completed. In other words, the CPU and GPU can work concurrently. With this insight, we can overlap the executions of the preparation step and the closing step in CPU with the executions of the GPU solvers.

\begin{figure}
\centering
\includegraphics[width=15cm]{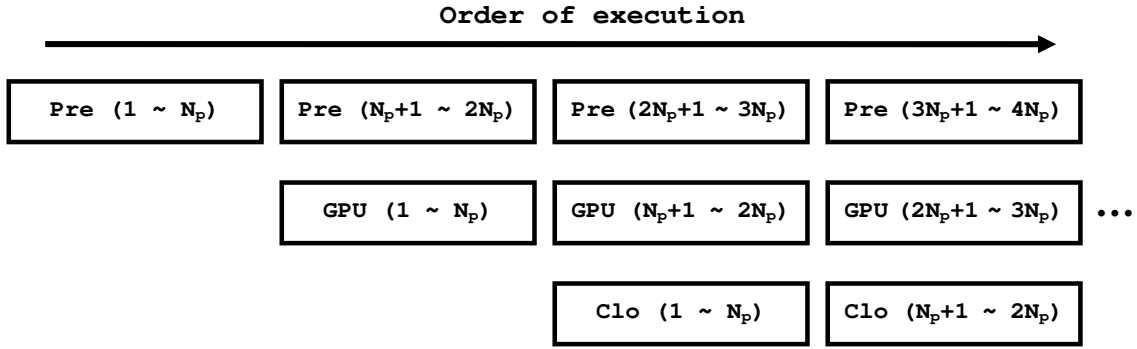}
\caption{Illustration of the concurrent execution between CPU and GPU. The abbreviations ``Pre'' and ``Clo'' stand for the preparation step and the closing step, respectively. The numbers in the parentheses indicate the indices of patch groups being calculated. The operations at the same column are performed concurrently. Accordingly, the preparation step and the closing step performed by CPU can be overlapped with the hydrodynamic solver or the Poisson solver performed by GPU.}
\label{fig:ConcurrentCPUGPU}
\end{figure}

As an illustration, let $N_p$ denote the number of patch groups calculated by one invocation of the GPU solver. For using $N_s$ stream objects, each of which contains the calculations of $N_g$ patch groups, we have $N_p=N_s\times N_g$. At the first step, the patch groups $1$ to $N_p$ are prepared by CPU. At the second step, the patch groups $1$ to $N_p$ are calculated by the GPU solver, while at the same time the patch groups $N_p\!+\!1$ to $2N_p$ are prepared by CPU. At the third step, the patch groups $N_p\!+\!1$ to $2N_p$ are calculated by the GPU solver, while at the same time the patch groups $2N_p\!+\!1$ to $3N_p$ are prepared by CPU, and the solutions of the patch groups $1$ to $N_p$ are copied to the corresponding pointers by CPU in the closing step. This procedure continues until the solutions of all targeted patch groups are obtained, as illustrated in Figure \ref{fig:ConcurrentCPUGPU}.

We note the similarity between the concurrent executions with the downstream memory copy, the kernel launch, and the upstream memory copy in the GPU solver (Fig. \ref{fig:AsyncMemcpy}) and the concurrent executions with the preparation step, the GPU solver, and the closing step (Fig. \ref{fig:ConcurrentCPUGPU}). In principle, having a larger $N_s$ can further decrease the communication time in the PCI Express bus. However, it will also deteriorate the efficiency of concurrent execution between CPU and GPU, because the amount of CPU workload that can be overlapped with GPU computing is reduced. Typically, we set $N_s=4$ to balance these two optimization approaches.

In order to test the efficiency of the concurrent execution between CPU and GPU, we execute each GPU solver in \emph{GAMER} and simultaneously perform a matrix summation in CPU. We keep the workload of the GPU solver fixed while varying the size of the matrix, and measure the total execution time in each case. Ideally, the total execution time should be equal to the one that consumes more time.

\begin{figure}
\centering
\includegraphics[width=15cm]{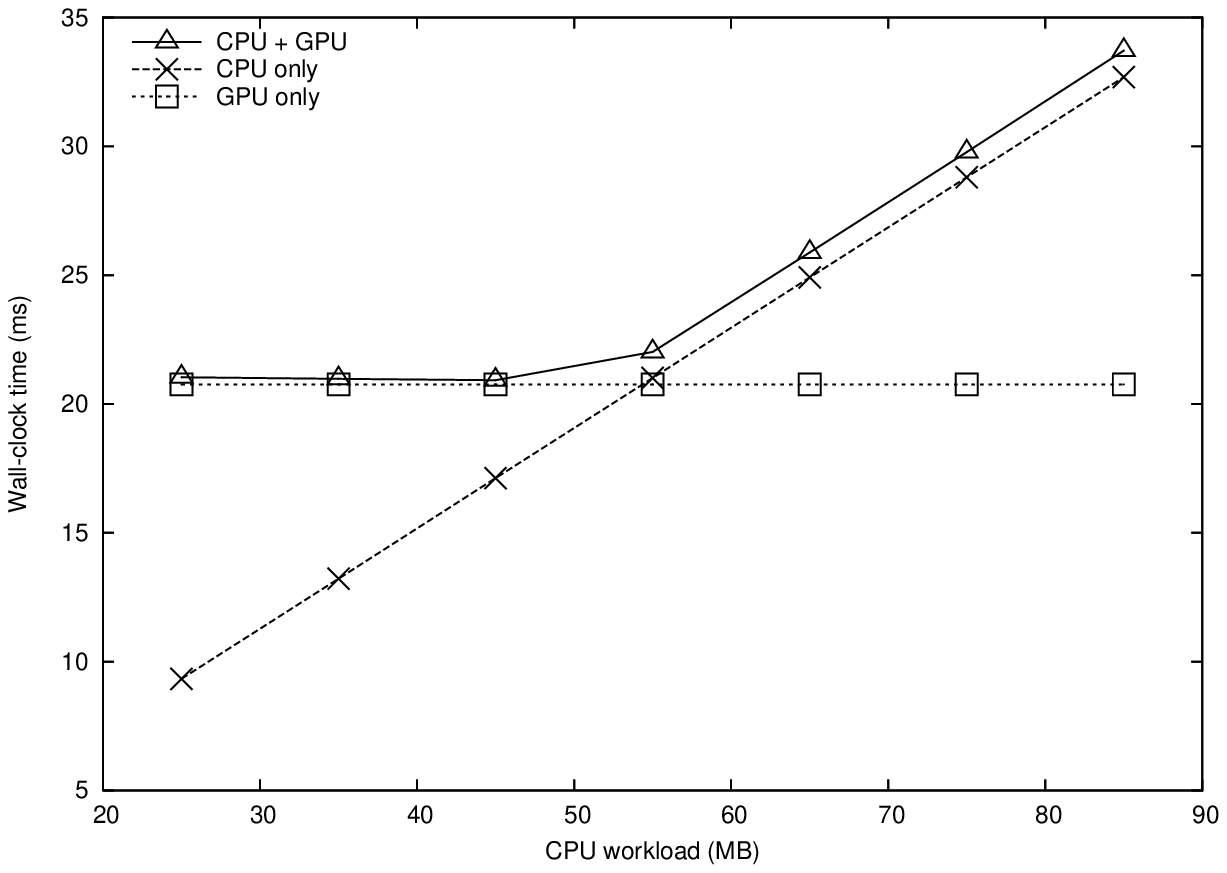}
\caption{Efficiency of the concurrent execution between CPU and GPU for the GPU hydrodynamic solver. The x-axis represents the size of the matrix summation performed by CPU.}
\label{fig:GPUFluidSolver_AsyncEfficiency}
\end{figure}

\begin{figure}
\centering
\includegraphics[width=15cm]{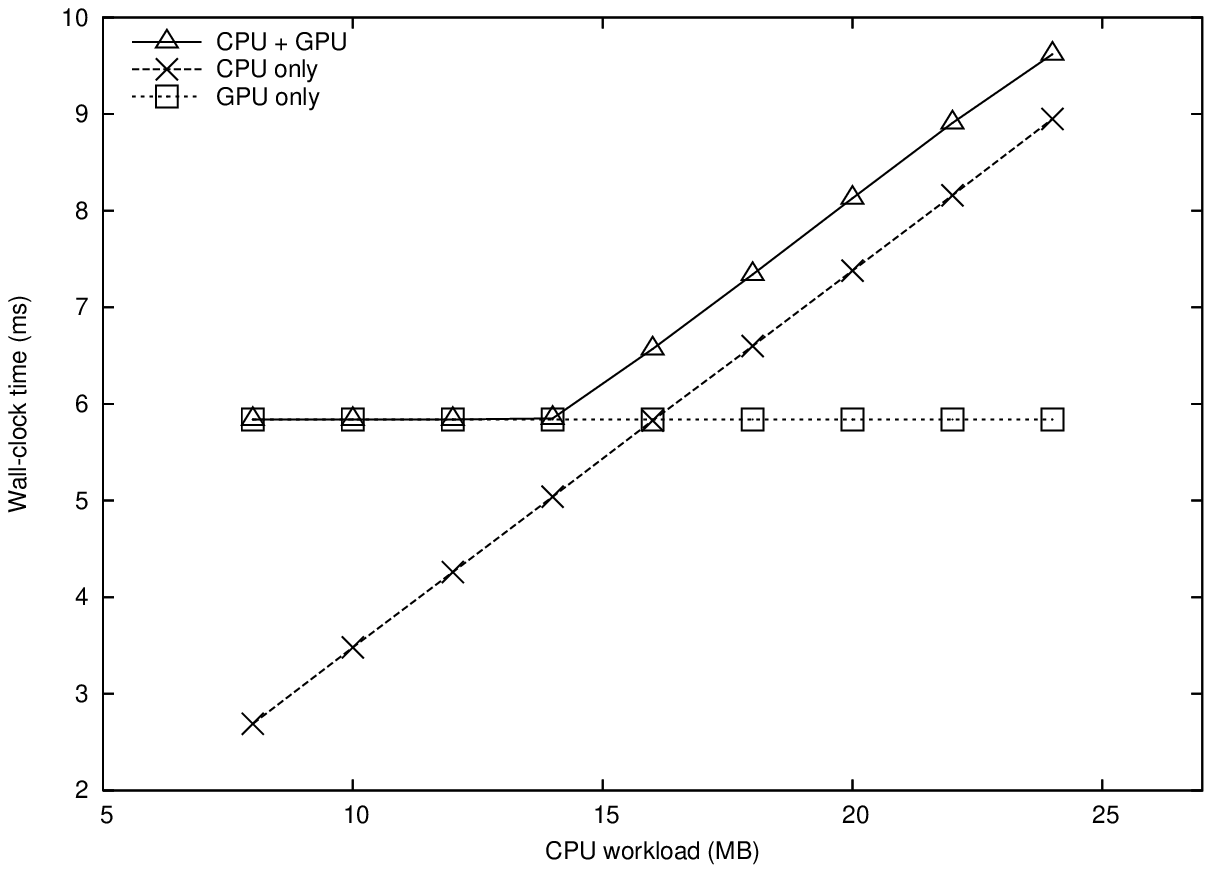}
\caption{Efficiency of the concurrent execution between CPU and GPU for the GPU Poisson solver. The x-axis represents the size of the matrix summation performed by CPU.}
\label{fig:GPUPoissonSolver_AsyncEfficiency}
\end{figure}

Figures {\ref{fig:GPUFluidSolver_AsyncEfficiency}} and {\ref{fig:GPUPoissonSolver_AsyncEfficiency}} show the test results for the GPU hydrodynamic solver and the GPU Poisson solver, respectively. The measurements are conducted in the NCHC system. As expected, in both solvers the total execution times are dominated by the tasks with higher workload, and good concurrent efficiencies are demonstrated. At the crossover point in Figure {\ref{fig:GPUFluidSolver_AsyncEfficiency}}, where the execution times of GPU and CPU are approximately equal, the measured wall-clock times of performing only the GPU hydrodynamic solver, only the matrix summation, and the concurrent execution between CPU and GPU are 20.76 ms, 21.01 ms, and 22.02 ms, respectively. The overhead is 4.8\%. At the crossover point in Figure {\ref{fig:GPUPoissonSolver_AsyncEfficiency}}, the measured wall-clock times of performing the GPU Poisson solver only, the matrix summation only, and the concurrent execution between CPU and GPU are 5.84 ms, 5.83 ms, and 6.57 ms, respectively. The overhead is 12.5\%. For both solvers, the overheads are less than 1.5\% when the GPU solvers dominate the execution times.

The performance improvement actually obtained by exploiting the concurrency between CPU and GPU is application-dependent, since the execution times of the preparation step and the closing step are related to the structure of domain refinement. We demonstrate the significance of utilizing this feature by comparing the performances in purely-baryonic cosmological simulations, which will be described in detail in \S\ 5.1.

\section{ACCURACY TESTS}

\emph{GAMER} is designed to achieve both high performance and high accuracy. All GPU kernels implemented in the code follow the same numerical algorithms as their CPU counterparts, except for certain optimized data managements which do not affect the accuracy. In other words, no numerical robustness is sacrificed for performance. In the following, we give the results of various standard tests to demonstrate the accuracy of the code. Single precision is assumed unless the floating-point precision is explicitly stated.

\subsection{Acoustic Wave Test}

As a first elementary test, we simulate the one-dimensional acoustic wave. This test is particularly useful for demonstrating the second-order accuracy of the hydrodynamic solver \citep{MC07a}. We construct the initial condition as following. The uniform background density and pressure are set to $\rho_0=1$ and $P_0=3/5$, and the ratio of the specific heats is set to $\gamma=5/3$. The sound speed is given by $c_s=(\gamma P_0/\rho_0)^{1/2}=1$. A sinusoidal density perturbation is adopted with a very small amplitude $\delta \rho/\rho_0=10^{-6}$ in order to avoid any non-linear effect (e.g., the wave steepening). The flow is initially at rest. Double precision is adopted to reduce the round-off errors. To verify the code accuracy in multi-dimensional cases, we choose the wave vector {\bf K} to be parallel to the diagonal of simulation box, i.e., ${\bf K=(1/\sqrt{3},1/\sqrt{3},1/\sqrt{3})}$.

\begin{figure}
\centering
\includegraphics[width=15cm]{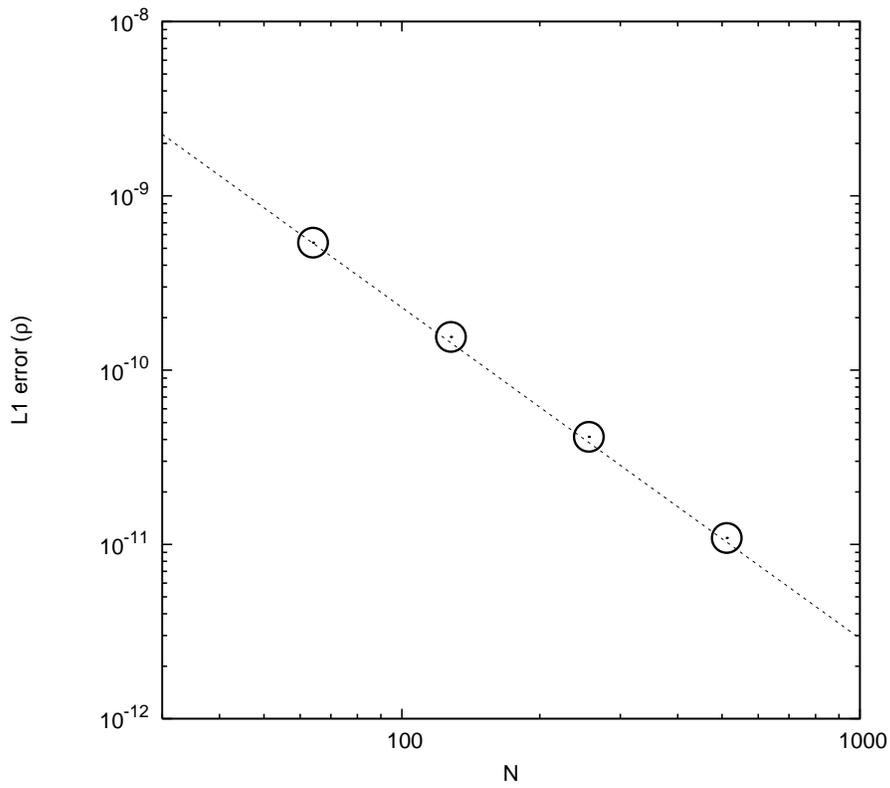}
\caption{Acoustic wave test. The wave vector is aligned with the diagonal of simulation box. The data points represent the L1 error norm of density as a function of numerical resolution (N).}
\label{fig:AcousticWave}
\end{figure}

We perform simulations with increasing resolutions: $N=64^3-512^3$, and estimate the L1 error norm of each case. The spatial coordinate is normalized to the grid size at $N=512^3$. In each case, the L1 error norm of density is defined as
\begin{equation}
L1(\rho)=\frac{1}{N}\sum\limits_{i}|\rho_i-\rho({\bf x}_i)|,
\end{equation}
where $\rho_i$ is the numerical solution of $i^{th}$ cell along the diagonal and $\rho({\bf x}_i)$ is the corresponding analytical solution. Figure \ref{fig:AcousticWave} shows the L1 error norm at $t=600$, as a function of spatial resolution. The error converges as $L1(\rho)\propto N^{-1.9}$, which is slightly slower than second order. Error distribution shows that error peaks sharply near local extreme values of fluxes, at which the hydrodynamic solver locally reduces to only first-order accurate. It is found that these errors may contaminate the convergence rate of numerical solutions. Nevertheless, the global convergence still approaches second order.

\subsection{Shock Tube Test}

The shock tube test \citep{Sod78} gives a great insight into the capability and accuracy of our hydrodynamic AMR program. Initially, the flow is stationary, with mass density and pressure jumps across a discontinuous interface. Afterwards, three characteristic waves with different propagation speeds are excited, namely, the rarefication wave, the shock wave, and the contact discontinuity. While the hydrodynamic quantities in the Euler equations are continuous across the rarefaction wave, the mass density is discontinuous across the contact discontinuity, and both the mass density, flow velocity, and pressure are discontinuous across the shock wave. A correct AMR scheme must fulfill the following criteria. The quantities to the left and the right of the shock front must satisfy the Rankine-Hugoniot shock jump conditions. All these features should be correctly captured by the simulation. Moreover, the domain refinement should be able to follow the wave propagations closely, the discontinuities should be resolved with a small number of cells, and there should be no spurious oscillations resulting from the domain refinement.

The initial conditions are constructed as following:
\begin{equation}
\rho_L=1, ~~~~~ \rho_R=0.125,
\label{eq:ShockTube_rho}
\end{equation}
\begin{equation}
P_L=1, ~~~~~ P_R=0.1,
\label{eq:ShockTube_Pres}
\end{equation}
\begin{equation}
u_L=u_R=0,
\label{eq:ShockTube_Vel}
\end{equation}
where the subscripts $L$ and $R$ indicate the states to the ``left'' and ``right'' of the initial discontinuous interface, respectively. The ratio of the specific heats $\gamma$ is set to $5/3$. The interface normal is chosen to be parallel to the x-axis. We set the size of the root level equal to 64, with 5 refinement levels ($\ell_{max}=5$). Accordingly, the effective resolution is 2048. The Dirichlet boundary condition is adopted to ensure no boundary effects to contaminate the wave propagations.

The refinement criterion is based on the density gradient. A cell is flagged for refinement if the relative change of density across the width of a single cell exceeds a given threshold $C_{\rho}$:
\begin{equation}
\frac{\triangle h_{\ell}}{\rho}\frac{\partial\rho}{\partial x}\ge C_{\rho}.
\label{eq:ShockTube_RefineCriterion}
\end{equation}
For the shock tube test, we have $C_{\rho}=0.03$. The size of the flag buffer ($N_b$) is set to 8.

\begin{figure}
\centering
\includegraphics[width=15cm]{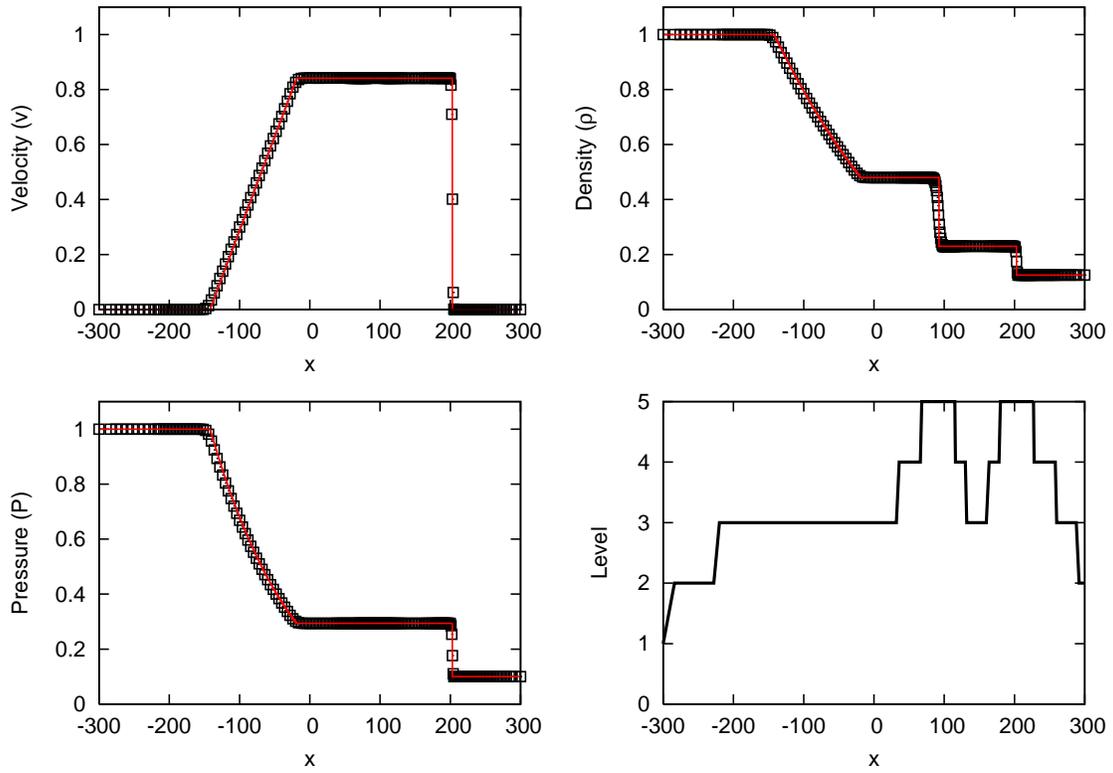}
\caption{Shock tube test. The top left, top right, bottom left, and bottom right panels show the velocity, mass density, pressure, and refinement level as a function of position at $t=110$, respectively. The initial discontinuous interface is placed at $x=0$. The solid lines depict the analytical solutions.}
\label{fig:ShockTube}
\end{figure}

Figure \ref{fig:ShockTube} shows the simulation results at $t=110$, along with the analytical solutions for comparison. The distance is normalized to the grid size at the maximum level. Both the contact discontinuity and the shock wave are refined to the maximum level, whereas the rarefication wave is refined to $\ell=3$. Only $2.34\%-7.81\%$ of the computational domain is refined to the maximum level during the entire simulation, and note that the discontinuous interface is refined to the maximum level from the beginning. The simulation results demonstrate a good resolution at discontinuities. The shock front is well resolved within three to four cells, and the contact discontinuity, which is potentially more diffusive, is resolved within seven to nine cells. We find a good agreement between the numerical solutions and the analytical solutions. The maximum numerical error is found to be about $1\%-3\%$ in the rarefication wave for both the mass density, velocity, and pressure. No unphysical oscillations are observed throughout the computational domain, and the flow velocity and pressure remain uniform across the contact discontinuity.

We adopt the individual-time-step scheme for the shock tube test. At $t=110$, it only takes 22 time steps to evolve the patches at the root level, whereas it takes $22\times 2^5=704$ steps to evolve the patches at the maximum level. A simulation with uniform time step is also performed for comparison, and no obvious differences in profiles are found between these two schemes.

\subsection{Sedov-Taylor Blast Wave Test}

The Sedov-Taylor blast wave test features a strong spherical shock and a self-similar flow. Initially, the mass density is uniform and the flow is at rest. The explosion is triggered by instantaneously releasing a point-like energy source, which is assumed to be much larger than the background thermal energy, into a homogeneous medium. Afterwards, a spherical shock propagates outward from the point of explosion. Since the background thermal energy is negligible compared to the explosion energy, the shock wave is strong. The challenge of the blast wave test lies in solving a spherically symmetric problem on the Cartesian grid. It may also verify the capability of \emph{GAMER} to accurately capture the propagation of strong shock waves.

A self-similar solution to the Sedov-Taylor blast wave is described in detail in \citet{LL87}. Let $\rho_1$ denote the initial background density, and $E_0$ be the total amount of explosive energy introducing at $t=0$. The explosion center is chosen to be the origin of the coordinate system. Then, from the dimensional analysis, the position and the propagation velocity of the shock front at $t>0$ are given by
\begin{equation}
r_s(t)=\beta\left(\frac{E_0t^2}{\rho_1}\right)^{1/5},
\label{eq:BlastWave_ShockFrontPos}
\end{equation}
\begin{equation}
v_s(t)=\frac{dr_s}{dt}=\frac{2\beta}{5}\left(\frac{E_0}{\rho_1t^3}\right)^{1/5},
\label{eq:BlastWave_ShockFrontVel}
\end{equation}
where $\beta$ is a constant depending on $\gamma$. For $\gamma=5/3$, $\beta\approx 1.15$. The values immediately behind the shock front can be derived by applying the Rankine-Hugoniot shock jump conditions in the strong shock limit, which gives
\begin{equation}
\rho_2=\left(\frac{\gamma+1}{\gamma-1}\right)\rho_1,
\label{eq:BlastWave_PostShockRho}
\end{equation}
\begin{equation}
v_2=\left(\frac{2}{\gamma+1}\right)v_s,
\label{eq:BlastWave_PostShockVel}
\end{equation}
\begin{equation}
P_2=\left(\frac{2}{\gamma+1}\right)\rho_1v_s^2.
\label{eq:BlastWave_PostShockPres}
\end{equation}
Finally, by introducing the dimensionless similarity variable
\begin{equation}
\xi=\frac{r}{r_s(t)}=\frac{r}{\beta}\left(\frac{\rho_1}{E_0t^2}\right)^{1/5},
\label{eq:BlastWave_SimilarityVariable}
\end{equation}
the Euler equations can be transformed into a set of ordinary differential equations, and the post-shock solutions can be obtained by direct integration.

We simulate the blast wave by initializing the background environment as $\rho_1$=1, $v_1=0$, and $E_1=10^{-5}$. The ratio of the specific heats $\gamma$ is set to $5/3$. A total thermal energy $E_0=8\times 10^5$ is injected into the eight central cells at the finest level. Note that since an averaging operation is performed during the initialization, the explosive energy is also injected into the eight central cells at coarser levels with an energy density one-eighth of that of their parent cells. The size of the root level is set to $64^3$ and the maximum level is chosen to be $\ell_{max}=5$, giving $2048^3$ effective resolution.

We adopt the pressure gradient as the refinement criterion so that the central region can be refined from the beginning. A cell is marked for refinement if the relative change of pressure across the width of a single cell exceeds a given threshold $C_P$:
\begin{equation}
\frac{\triangle h_{\ell}}{P}\frac{\partial P}{\partial x}\ge C_{P}.
\label{eq:BlastWave_RefineCriterion}
\end{equation}
For the blast wave test, we set $C_P=1.0$. The size of the flag buffer ($N_b$) is chosen to be 8. Since the background quantities are continuous across the simulation box, the periodic boundary condition is adopted.

\begin{figure}
\centering
\includegraphics[width=15cm]{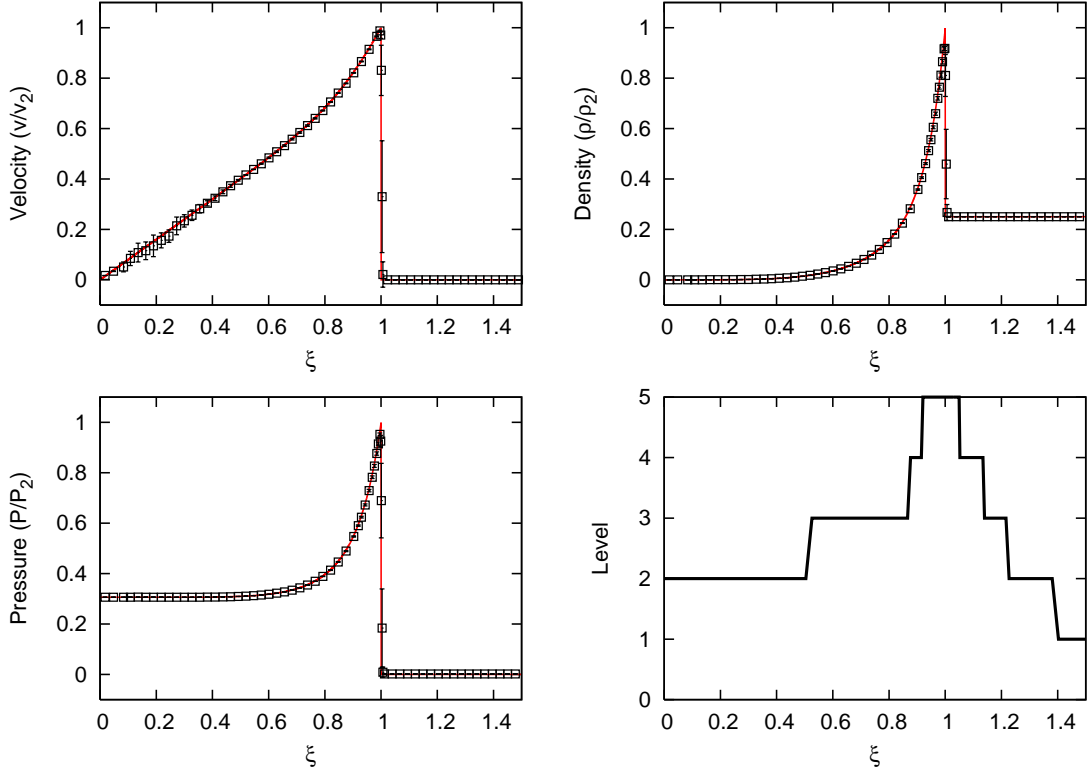}
\caption{Sedov-Taylor blast wave test. The top left, top right, bottom left, and bottom right panels show the shell-averaged velocity, mass density, pressure, and refinement level as a function of radius at $t=2000$, respectively. The error bars represent the standard deviations to the average values. The hydrodynamic variables are normalized to the values directly behind the shock front, and the radius is normalized to the position of the shock front. The solid lines depict the analytical solutions. In the right bottom panel, the refinement level is recorded along the x-axis. }
\label{fig:BlastWave}
\end{figure}

The simulation results at $t=2000$ are shown in Figure \ref{fig:BlastWave}, along with the analytical solutions for comparison. The hydrodynamic variables are normalized to the values directly behind the shock front (Eqs. [\ref{eq:BlastWave_PostShockRho}]-[\ref{eq:BlastWave_PostShockPres}]), and the dimensionless radius ($\xi$) is adopted for the spatial coordinate. The numerical results shown here are the spatial averages over radial bins with thickness equal to the resolution at the finest level. The standard deviations (asphericity) from the shell-averaged values in each bin are represented by error bars. We find a good agreement between the numerical solutions and the analytical solutions. The propagation of the strong shock front is accurately captured, with a compression ratio $\rho(\xi=0.998)/\rho_1=3.67$. The shock front is well resolved within three cells. Moreover, despite using the Cartesian geometry to solve a spherically symmetric problem, the standard deviations from the shell-averaged values are small. The maximum deviation is found in the velocity profile at the radius $\xi=0.1\sim0.3$, where the effective resolution is only $256^3$.

The distribution of refinement levels along the x-axis is also shown in Figure \ref{fig:BlastWave} for illustration. Only the region surrounding the shock front is refined to the maximum level, which is 1.38\% of the entire simulation box. Due to the proper-nesting constraint, the refinement levels gradually downgrade to $\ell=0$ and $\ell=2$ in the upstream and downstream regions, respectively. We adopt the individual time step for the blast wave test. At $t=2000$, only 347 time steps are required for the root level, while 11,104 time steps are required for the finest level.

\subsection{Point-Masses Poisson Solver Test}

In order to verify the accuracy of the multi-level Poisson solver described in \S\ 2.3, we perform tests to measure the gravitational accelerations between point masses. Two unit-mass particles are randomly placed in a simulation box with $32^3$ root-level cells. The maximum refinement levels are set to $\ell_{max}=0-6$, giving $32^3-2048^3$ effective resolutions. The refinement criterion is set to $\rho_{\ell}>0$, so that the eight cells surrounding each particle are always refined to the maximum level. At each level, we first compute the mass density using the cloud-in-cell (CIC) interpolation scheme \citep{HE81}. Afterwards, the gravitational potential and acceleration at each cell are evaluated, and the acceleration of each particle is obtained by the inverse CIC interpolation. Ideally, with an increasing refinement level, the force resolution should be improved and approach the cell size of the current refinement level.

\begin{figure}
\centering
\includegraphics[width=15cm]{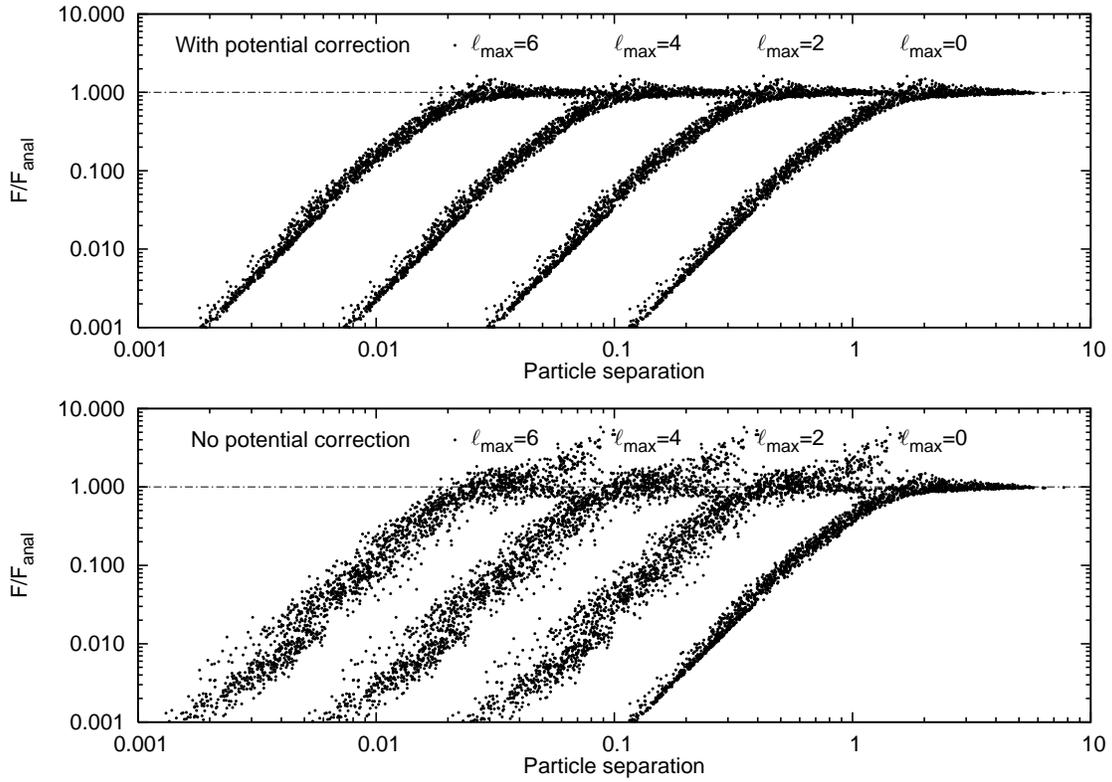}
\caption{Poisson solver test. Two particles are randomly distributed in the simulation box and the gravitational acceleration is evaluated using the multi-level Poisson solver. Here we show the numerical results divided by the analytical solutions, using 0 to 6 refinement levels. The particle separation is normalized to the root-level grid size. The top and bottom panels show the results with and without potential correction, respectively.}
\label{fig:PointMasses_Poisson}
\end{figure}

Figure \ref{fig:PointMasses_Poisson} shows the evaluated potentials divided by the analytical solutions. The contributions from image particles due to the periodic boundary condition are calculated using the Ewald summation technique \citep{Hernquist91}. For comparison, we also show the results without applying any potential correction. These two cases should give identical results when only the root level is used and the multi-level Poisson solver reduces to a pure FFT solver. For $\ell_{max}>0$, we see that the results without potential correction significantly deviate from the analytical solutions. It reveals the fact that the interpolation errors at coarse-fine boundaries can seriously contaminate the solutions at finer levels. On the contrary, when the potential correction is applied, the force resolution can approach the cell size at the given maximum level. The effective force resolution remains approximately two cells in each case, and the scatters of solutions are highly suppressed. Note that the small solution scatters arise primarily from the fictitious anisotropic force of the square meshes. We also notice that the error spatial distributions in the cases $\ell_{max}>0$ are similar to that in the case $\ell_{max}=0$, indicating that the numerical errors are dominated by the CIC interpolation.

To provide a more quantitative analysis, we further compare in detail the solutions obtained from the hierarchical AMR grids with the uniform-grid solutions, and investigate the numerical errors introduced from the spatial interpolations at the coarse-fine boundaries after the potential correction. One of the approaches to verify the solution accuracy is to fix the particle distributions, compare the numerical solutions to the analytical solutions using different numbers of refinement levels, and estimate the solution convergence rate. However, this approach does not provide a good insight into the convergence behavior for the cases with particle separations close to the minimum cell size, because the errors will be dominated by the CIC interpolation when using less refinement levels. For example, for a particle pair with separation equal to 2 finest cells at $\ell_{max}=6$, the same particle separation only corresponds to 0.5 finest cell when using $\ell_{max}=4$, in which the numerical solution deviates significantly from the analytical solution due to the softening effect of the CIC method. Accordingly, it is inappropriate to compare the force solution of this particle pair using refined meshes of different levels. To resolve this issue and quantify the numerical errors for the cases that particle separations are close to the given minimum cell size, we proceed the analysis as following.

First, for each particle pair, we fix its relative orientation, but normalize the particle separation by the current minimum cell size when adopting different maximum refinement levels. The multi-level Poisson solver is then applied to evaluate the gravitational force between each particle pair, and the forces introduced by image particles are subtracted. By doing so, since the minimum cell size provides the only length scale, the numerical solution obtained using a higher refinement level can be easily rescaled and compared with the root-level solution. Ideally, these two solutions should be identical, if the errors introduced from the coarse-fine boundaries are negligible. For example, for a given particle pair, the gravitational force obtained using $\ell_{max}=1$ should be exactly four times larger than that obtained using $\ell_{max}=0$.

\begin{figure}
\centering
\includegraphics[width=15cm]{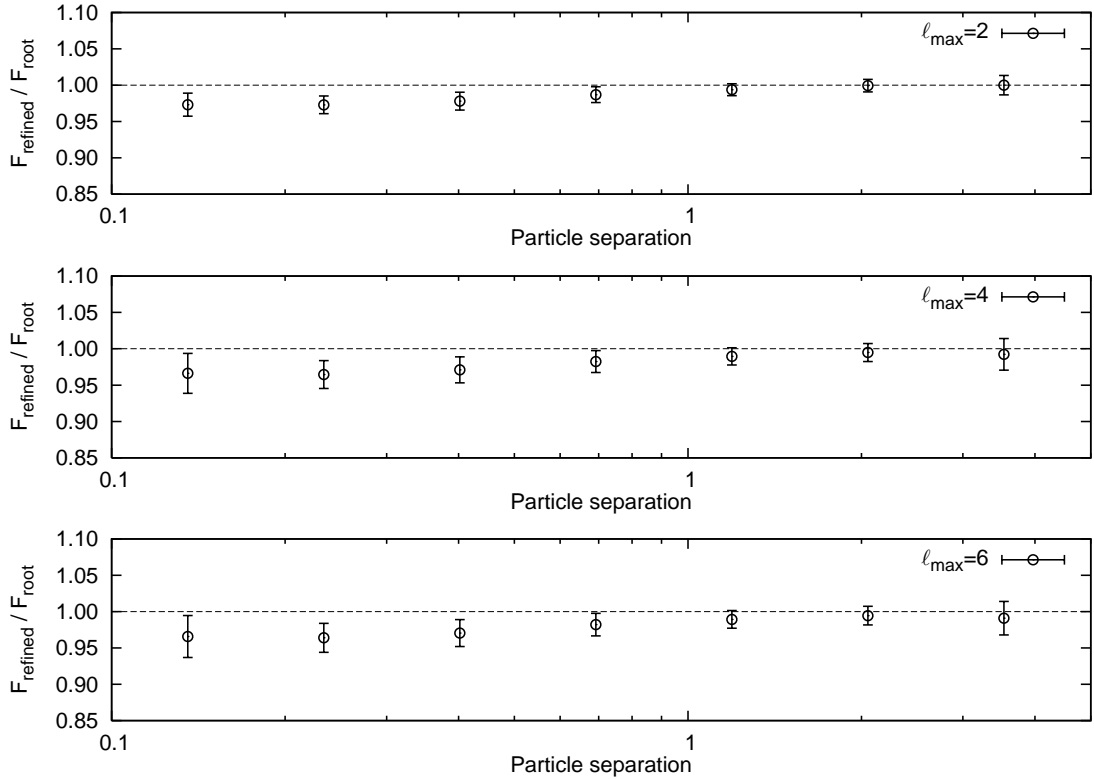}
\caption{Poisson solver test. Here we plot the scaled refinement-level solutions divided by the root-level solutions, using 2, 4, and 6 refinement levels. The error bars represent the standard deviations at different data bins. The particle separation is normalized to the minimum grid size in each case. The dashed lines represent the ideal results, in which the scaled refinement-level solutions are equal to the root-level solutions.}
\label{fig:PointMasses_Poisson_RMS}
\end{figure}

Figure \ref{fig:PointMasses_Poisson_RMS} plots the ratio between the rescaled solutions at refinement levels and the solutions obtained using only the root level. Data are divided into bins that are equally spaced in logarithmic scale, and the standard deviation in each data bin is represented by an error bar. For the case $\ell_{max}=6$, the average ratios at different data bins range from $96.4\%$ to $99.4\%$, and the maximum RMS value is $2.89\times 10^{-2}$. For particle separations larger than one cell, the average ratios are above $99.0\%$. Also note that the errors for $\ell_{max}=4$ and $\ell_{max}=6$ are nearly indistinguishable. It verifies that the errors introduced from the coarse-fine boundaries are suppressed after a few levels of refinement, and the accuracy of the point-masses Poisson solver test is always dominated by the CIC induced errors at the finest level. In other words, it demonstrates that, within a fixed small error, the multi-level Poisson solver is able to give a force resolution equivalent to a uniform-mesh PM solver whose cell size is equal to that in the current maximum refinement level, regardless of the level of refinement.

\subsection{Jean's Instability Test}

To further test the accuracy of Poisson solver and give a quantitative analysis of the integration scheme in a hydrodynamic + self-gravity system, we test the problem of Jean's instability. For a small-amplitude perturbation, Eqs. (\ref{eq:MassConserve})-(\ref{eq:EnergyConserve}) and (\ref{eq:Poisson}) can be linearized and the dispersion relation is given by
\begin{equation}
\omega^2=c_s^2[k^2-k_J^2],
\label{eq:Jeans_Dispersion}
\end{equation}
where $k_J$ is the Jean's wave vector
\begin{equation}
k_J^2=\frac{4\pi G\rho_0}{c_s^2},
\label{eq:Jeans_WaveVector}
\end{equation}
$c_s=(\gamma P_0/\rho_0)^{1/2}$ is the sound speed, and $\rho_0$ and $P_0$ are the background density and pressure, respectively. For $k<k_J$, a growing-mode solution can be found and the density perturbation grows exponentially as
\begin{equation}
\rho(t)=\rho_0+\delta\rho\sin(kx)e^{\sqrt{-\omega^2}t},
\label{eq:Jeans_Density}
\end{equation}
where $\delta\rho$ is the amplitude of initial density perturbation.

\begin{figure}
\centering
\includegraphics[width=15cm]{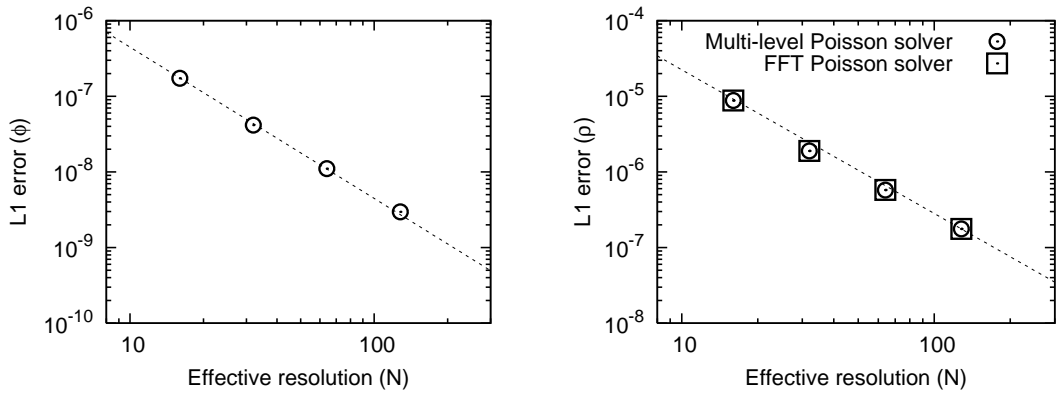}
\caption{Jean's instability test. The left panel shows the L1 error norms of the potential. The right panel shows the L1 error norms of density obtained by the multi-level Poisson solver (circles) and the FFT Poisson solver (squares). Two schemes give nearly identical results.}
\label{fig:JeansInstability}
\end{figure}

We simulate the Jean's instability problem by having $\delta\rho=10^{-6}$, $G=10^{-3}$, $\rho_0=1$, $P_0=3/5$, and $\gamma=5/3$. The wave length is chosen to be equal to the size of simulation box, which has a $16^3$ root level with zero to three refinement levels. The effective resolutions are therefore $N=16^3-128^3$. Double precision is adopted to reduce the round-off errors. Since the sinusoidal density distribution has an analytical solution to the Poisson equation, we first compare it to the numerical solutions. Figure \ref{fig:JeansInstability} shows the L1 error norm of potential. Encouragingly, the second-order convergence is verified. We then perform simulations until the perturbation amplitudes have grown two orders of magnitude, and compare the numerical results to the analytical predictions. The L1 error norm of density is shown in Figure \ref{fig:JeansInstability}. The error converges as $L1(\rho)\propto N^{-1.9}$, which is consistent with the convergence rate observed in the acoustic wave test. For comparison, we also conduct simulations using only the FFT Poisson solver with the same effective resolutions. As shown in Figure \ref{fig:JeansInstability}, the numerical errors observed in these two schemes are nearly identical.

\subsection{Spherical Collapse Test}

In the spherical collapse test, an initial overdense perturbation with spherical symmetry is placed in the Einstein-de Sitter space ($\Omega_m=1$). The background pressure is assumed to be negligible. Since the system is gravitational bound, the overdense region first expands with the Hubble flow, and eventually begins to collapse at the ``turn around'' time. The mass shells at larger radii also collapse at later times, and a strong shock wave propagates outward. Since the turnaround radius provides the only length scale in this problem, the solution is self similar. The spherical collapse test is the most comprehensive test for the three-dimensional hydrodynamic code with self gravity. It simultaneously involves several essential properties of the code, namely, the accuracy of both hydrodynamic and Poisson solvers, the shock-capturing capability, the domain refinement due to the density contrast, and the accuracy of solving a spherically symmetric problem using the Cartesian grid.

An analytical solution to the spherical collapse problem was given by \citet{Bertschinger85}, for both collisional and collisionless components. Here we only consider the collisional case. Let $\delta_i$ and $R_i$ denote the density contrast and the radius of the top-hat overdense perturbation at an initial time $t_i$. The turnaround radius, from where the mass shells cease expanding and start to collapse, is given by
\begin{equation}
r_{{\rm ta}}(t)=\left(\frac{4}{3\pi}\frac{t}{t_i}\right)^{8/9}\delta_i^{1/3}R_i.
\label{eq:SC_TurnaroundRadius}
\end{equation}
Accordingly, the dimensionless radius and the fluid variables are defined as
\begin{equation}
\lambda\equiv\frac{r}{r_{{\rm ta}}},
\label{eq:SC_DimensionlessRadius}
\end{equation}
\begin{equation}
\tilde{\rho}(\lambda)\equiv\frac{\rho}{\rho_b},
\label{eq:SC_DimensionlessRho}
\end{equation}
\begin{equation}
\tilde{v}(\lambda)\equiv v\left(\frac{t}{r_{{\rm ta}}}\right),
\label{eq:SC_DimensionlessVel}
\end{equation}
\begin{equation}
\tilde{P}(\lambda)\equiv\left(\frac{P}{\rho_b}\right)\left(\frac{t}{r_{{\rm ta}}}\right)^2.
\label{eq:SC_DimensionlessPre}
\end{equation}
From Eqs. (\ref{eq:SC_DimensionlessRadius})-(\ref{eq:SC_DimensionlessPre}), the Euler equations can be reduced into a set of ordinary differential equations, and the analytical solutions can be calculated by direct integration.

We perform the spherical collapse test by setting $\delta_i=0.01$ and $R_i=160$ at $a=10^{-5}$ in a comoving box, in which the periodic boundary condition is adopted. The ratio of the specific heats $\gamma$ is chosen to be $5/3$, and the background pressure is set to an arbitrarily low value. Since the comoving coordinates are adopted in this test, the initial velocity is set to zero, which corresponds to an unperturbed Hubble flow in the physical coordinates. The size of the root level is set to $128^3$ and the maximum refinement level is four, giving $2048^3$ effective resolution. Note that the radius of the top-hat density distribution is much smaller than the size of the simulation box, so that the image density introduced by the periodic boundary condition should not severely infect the numerical results, and the overdense perturbation can be regarded as an isolated system.

We adopt the amplitude of density as the only refinement criterion. Any cell at level $\ell$ is marked for refinement if it exceeds the density threshold: $\rho_{\ell}>8^{\ell}$. Accordingly, the overdense region is refined to $\ell=1$ at the beginning. The size of the flag buffer ($N_b$) is chosen to be 8.

\begin{figure}
\centering
\includegraphics[width=15cm]{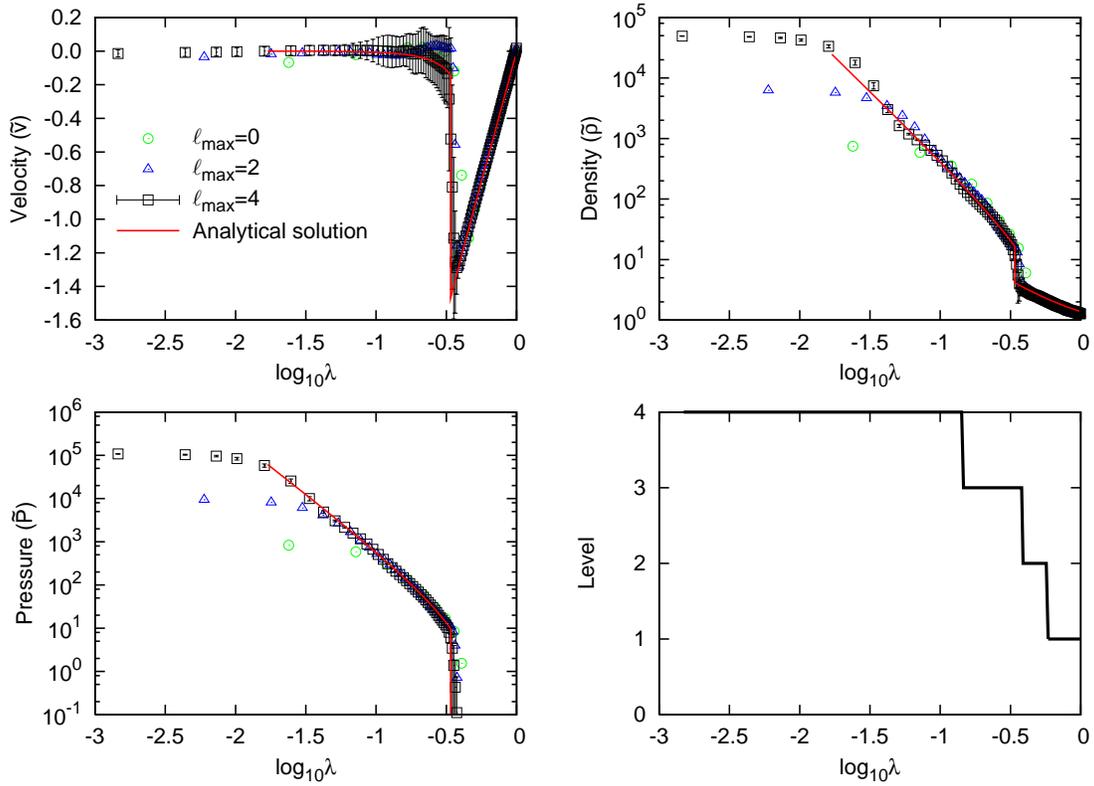}
\caption{Spherical collapse test. The top left, top right, bottom left, and bottom right panels show the shell-averaged dimensionless velocity, mass density, pressure, and refinement level as a function of the dimensionless radius at $a=0.09$, respectively. The circles, triangles, and squares show the numerical solutions using zero, two, and four refinement levels, respectively. The error bars represent the standard deviations to the shell-averaged values for the run using four refinement levels. The solid lines depict the analytical solutions. In the right bottom panel, the refinement level is recorded along the x-axis.}
\label{fig:SphericalCollapse}
\end{figure}

The simulation results at $a=0.09$ is shown in Figure \ref{fig:SphericalCollapse}. We plot the shell averages of the dimensionless variables defined in Eqs. (\ref{eq:SC_DimensionlessRadius})-(\ref{eq:SC_DimensionlessPre}). The analytical solutions derived by \citet{Bertschinger85} are also depicted for comparison. In order to test the convergence of the numerical solutions, we perform simulations with zero, two, and four refinement levels, giving $128^3$, $512^3$, and $2048^3$ effective resolutions, respectively. A good agreement between the numerical and analytical solutions is found. The simulation run with a higher maximum refinement level indeed probes the solutions in a more central region, where only two to four cells in the most central region are incapable to follow the rising density and pressure profiles predicted by the analytical solutions. This is due to the second-order TVD scheme that requires 3-point interpolation and has the tendency to smooth out the local extreme values. The propagation and the jump conditions of the strong spherical shock are accurately captured.

In order to test the property of spherical symmetry, we also show the standard deviations from the shell-averaged values for the simulation run with four refinement levels. No significant deviations are observed in both the density and the pressure profiles, whereas a relative large deviation is found in the velocity profile in the postshock region where the infall velocity abruptly drops to zero. It is consistent with the result given by \citet{Teyssier02}. Nevertheless, the shell-averaged velocity profile agrees well with the analytical solution even in the postshock region.

The result of domain refinement along the x-axis for the simulation run with four refinement levels is also shown in Figure \ref{fig:SphericalCollapse}. Note that neither the density gradient nor the pressure gradient are adopted as the refinement criteria in this test, therefore, the shock front is not necessarily refined to the maximum level. At $a=0.09$, only 0.01\% region is refined to the maximum level. Due to the self-similar property, the volume of the overdense region increases as the shock propagates outward, and hence the refined region increases as well. However, still only 0.02\% region is refined to the maximum level at $a=0.3$.

\section{PERFORMANCE TESTS}

The timing measurements described in \S\ 3 mainly focus on the performances of the hydrodynamic and Poisson solvers. Although there is no doubt that these two solvers are the most time-consuming parts in \emph{GAMER}, it is still not clear whether other parts of the code will become the performance bottlenecks, especially when the computation times of both solvers are highly reduced by using GPUs. For example, the ghost-zone preparation, the domain refinement, and the network bandwidth may also impact the performance. Accordingly, in this section, we perform detailed timing analyses for different parts of the code as well as the overall performance.

The accuracy tests described in \S\ 4 are inadequate for a persuasive timing analysis, since the profiles of these solutions are too simple compared to realistic astrophysical simulations. Therefore, to have a more comprehensive timing analysis, we measure the performance in purely-baryonic cosmological simulations. The initial condition is constructed by using CMBFAST \citep{SZ96} on $256^3$ grids in a $\Lambda$CDM universe at redshift $z=99$. The cosmological parameters are chosen to be $\Omega_m=0.3$ and $\Omega_{\Lambda}=0.7$, and the size of the periodic comoving box is set to 100 $h^{-1}$Mpc. In the following, we first describe the performance of \emph{GAMER} in a single-GPU system, and then follows the multi-GPU performance.

\subsection{Single-GPU Performance}

To fit into the CPU memory of a single node, we set the size of the root level equal to $128^3$. The initial condition is obtained by downgrading the $256^3$ initial condition in order to have consistent results among runs with different spatial resolutions. Five refinement levels are adopted to give $4096^3$ effective resolution. The corresponding spatial resolution is therefore 24 $h^{-1}$Kpc. We rebuild the refinement map every four steps, which provides the balance between performance and accuracy. Accordingly, the flag buffer size is set to $N_b=4$, so as to prevent the information in the flagged cells from propagating out of the refined regions before the refinement map is rebuilt.

The refinement criterion is similar to the one adopted in the spherical collapse test described in \S\ 4.6. A cell is labeled as refinable if its local density exceeds a level-dependent threshold: $\rho_{\ell}\ge8^{\ell+1}\rho_b$. This kind of ``quasi-Lagrangian'' refinement strategy is often adopted in the AMR cosmological simulations. It roughly fixes the total mass within each cell at different refinement levels. Moreover, when the dark matter particles are included and the standard particle-mesh method is used to calculate the gravitational potential, this refinement strategy naturally provides an adaptive soften length according to the level-dependent grid size, thereby minimizing the effect of two-body relaxation.

\begin{figure}
\centering
\includegraphics[width=15cm]{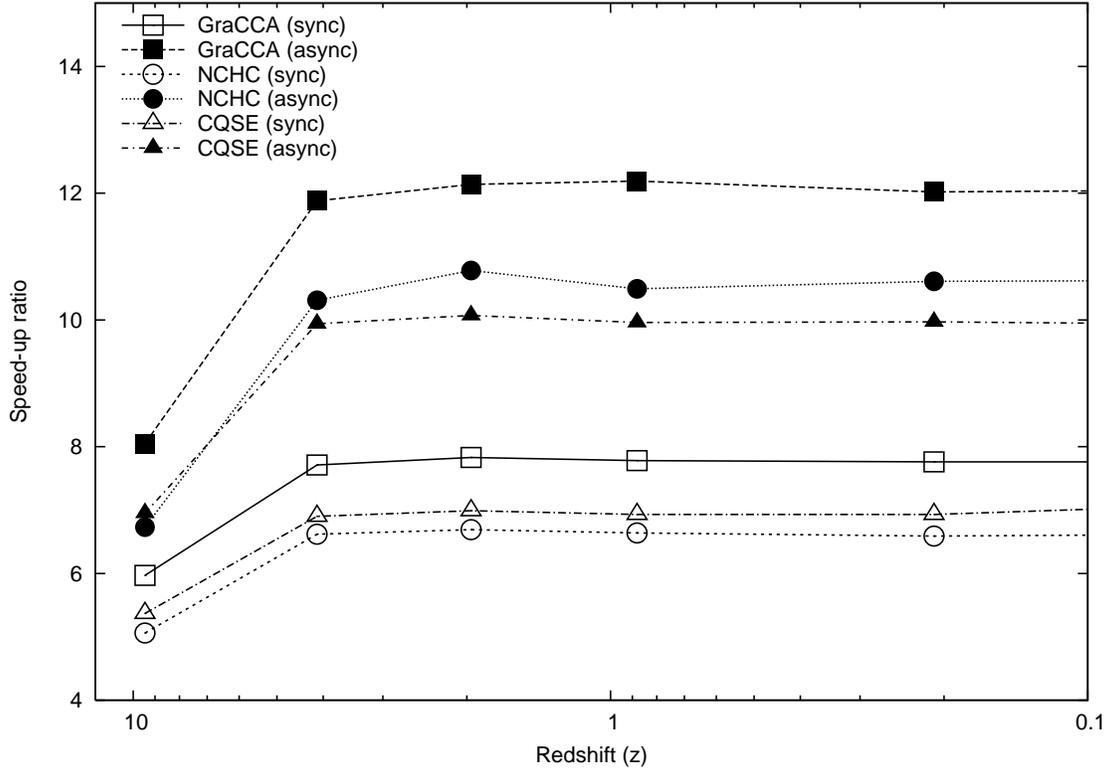}
\caption{Performance speed-up ratios as a function of redshift in different hardware implementations. We measure the performance in purely-baryonic cosmological simulations. The speed-up ratios are measured by comparing the runs with single-GPU acceleration to the runs using singe CPU only. The abbreviations ``async'' and ``sync'' represent the timing results with and without the concurrent execution between CPU and GPU, respectively. The sizes of the root levels are set to $128^3$ and the maximum refinement levels are set to 5 in all tests.}
\label{fig:GPUvsCPU_vs_Redshift__NRank1}
\end{figure}

Figure \ref{fig:GPUvsCPU_vs_Redshift__NRank1} shows the speed-up ratio of one GPU over one CPU as a function of the redshift. We measure the performances at six different redshifts: $z=9.45,~4.12,~1.96,~0.88,~0.21,~0$. To provide more robust results, we perform the timing analyses in the GraCCA system, NCHC, and CQSE, respectively. In each system, we further compare the performances with and without activating the concurrent execution between CPU and GPU described in \S\ 3.5. It shows that, as the concurrent execution is enabled, an order of magnitude performance improvements are demonstrated in all three different hardware implementations. The maximum sustained speed-up ratios are 12.19, 10.78, and 10.07 in the GraCCA system, NCHC, and CQSE, respectively, and the speed-up ratios are approximately constant when $z\le4.12$. It demonstrates the significant performance improvement by using GPU.

We notice that, at $z=9.45$, the performance speed-up ratios drop about 40\% compared to the maximum sustained performance in each GPU system. This is not surprising since at that time massive halos are not yet formed and thus the total number of patches at refined levels is relatively small. Consequently, the calculation time of the Poisson solver is dominated by the root-level FFT, which is computed by CPU. Nevertheless, factors of 8.05, 6.73, and 6.95 performance speed-ups are still achieved in the three systems at $z=9.45$.

In Figure \ref{fig:GPUvsCPU_vs_Redshift__NRank1}, we see considerable performance deterioration as the concurrent execution is disabled. The maximum sustained speed-up ratios drop to 7.83, 6.69, and 7.43 in the GraCCA system, NCHC, and CQSE, respectively. It reveals the large computational overheads associated with the preparation steps and the closing steps performed by CPU, for both the hydrodynamic and Poisson solvers. This result is reasonable since in the cosmological simulation, structures form hierarchically, and the number of grids used to resolve a substructure at a given refinement level is generally on the order of $10^3$. In \emph{GAMER}, it approximately corresponds to the size of a single patch group. Consequently, nearly all patch groups require spatial interpolations to calculate the ghost-zone values, and it results in a computation time comparable to or even longer than the execution time of each GPU solver. Therefore, it is necessary to simultaneously utilize both the computational power of CPU and GPU, thereby hiding the computation times of the preparation steps and the closing steps by the executions of the GPU solvers. For the cases where the CPU workload is higher, the execution time in GPU can also be hidden behind the CPU computation. The considerable performance improvements as the concurrent execution is activated also verify the good concurrent efficiencies as shown in Figures \ref{fig:GPUFluidSolver_AsyncEfficiency} and \ref{fig:GPUPoissonSolver_AsyncEfficiency}.

Table \ref{table:TimingGAMER} shows the timing results of different parts in \emph{GAMER} at $z=0.88$. We perform the timing measurements in three different hardware implementations, each of which includes the results using GPU with and without the concurrent execution, and the results using CPU only. First we notice that, even when the concurrent execution is enabled, the speed-up ratios of both the hydrodynamic and Poisson solvers are still lower than the results given in Tables \ref{table:TimingHydro} and \ref{table:TimingPoisson}. It implies that the computational overheads associated with the preparation step and the closing step of both solvers dominate the computation time. In other words, these computational overheads are the performance bottlenecks in the current version of \emph{GAMER}, and therefore any further optimization of the GPU solvers alone will not improve the overall performance at all. However, we must emphasize that the three-dimensional relaxing TVD scheme described in \S\ 2.2 has a relatively low arithmetic intensity, compared to other high-order shock-capturing schemes, for example, the third-order PPM method. Therefore, it is promising to further improve both accuracy and performance by implementing an alternative higher-order hydrodynamic scheme in the code.

In Table \ref{table:TimingGAMER}, we see that the overall performance is still dominated by the two solvers. The computation times of the gravitational acceleration, the data copies for the buffer patches, and the domain refinement are about 8\%, 2\%, and 1\% of the total simulation time. The calculation time of the CFL condition is negligible. Note that in the CQSE and NCHC systems, implementing the evaluation of the gravitational acceleration into GPU turns out to provide lower performances than using CPU. It reveals the trickiness of using GPU, where poor performance may be obtained if the arithmetic intensity of the GPU kernel is not high enough and hence the performance degrades due to the additional communication time in the PCI Express bus. Also note that here the preparation of the buffer-patch data do not involve any network communication. Finally, since in these tests we reconstruct the domain refinement in every four steps, the timing results of the refinement operations recorded here are the average values in order to compare with the computation times of other operations.

\begin{figure}
\centering
\includegraphics[width=15cm]{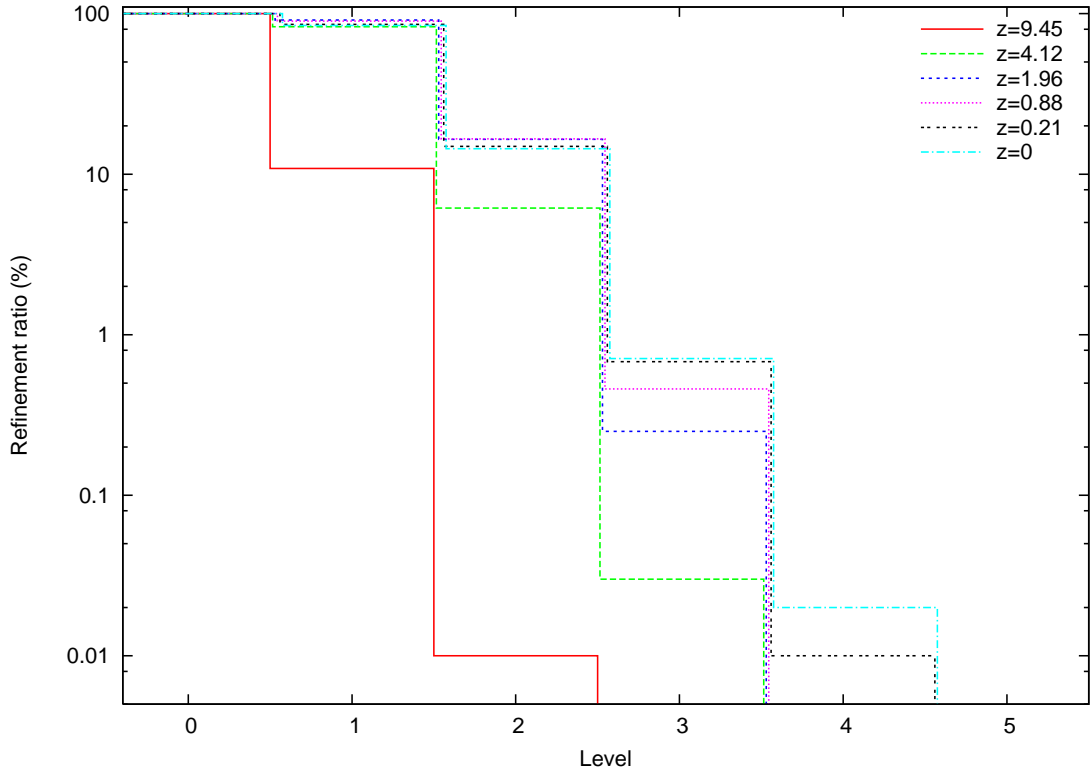}
\caption{Domain refinement ratios as a function of refinement levels in a purely-baryonic cosmological simulation. We show the results at six different redshifts. The lines of the refinement ratios at different redshifts are slightly shifted in the horizontal direction in order to be visually distinguishable.}
\label{fig:RefinementRatio_vs_Redshift}
\end{figure}

Figure \ref{fig:RefinementRatio_vs_Redshift} shows the refinement ratios at different redshifts as a function of the refinement level. The refinement ratio is defined as the ratio of the volume of space refined to a given level to the volume of the entire simulation box. At $z=9.45$, only 10.86\% and 0.01\% of the simulation box are refined to the level one and two, respectively. It verifies that the performance drop at $z=9.45$ shown in Figure \ref{fig:GPUvsCPU_vs_Redshift__NRank1} is due to the lack of patches at refined levels. In comparison, when there are sufficient number of patches at higher levels after $z\le4.12$, the performance approaches the maximum sustained speed-up ratio. Moreover, note that the total computation time of the simulation is dominated by the evolution at lower redshifts, when lots of substructures formed and, correspondingly, the total number of patches at refined levels is high enough for exploiting the computational power of GPU. Timing experiment shows that the accumulated computation time before $z>4.12$ is less than 0.03\% of the total computation time over $0\le z\le99$. Therefore, the overall performance improvement during the entire simulation does not suffer from the relatively low performance at higher redshifts. Totally 2608 steps are required to reach $z=0$, and it took only 16 hours of wall-clock time by using only one GPU and one CPU of the GraCCA system.

\subsection{Multi-GPU Performance}

Utilizing multiple GPUs across different nodes requires the network communication. Since the computation time is highly reduced by an order of magnitude as demonstrated in the previous subsection, correspondingly, the communication time becomes more critical. The data transfer time may easily become the performance bottleneck if it is not properly optimized. Therefore, in this section, we investigate the importance of network communication in \emph{GAMER}.

\begin{figure}
\centering
\includegraphics[width=15cm]{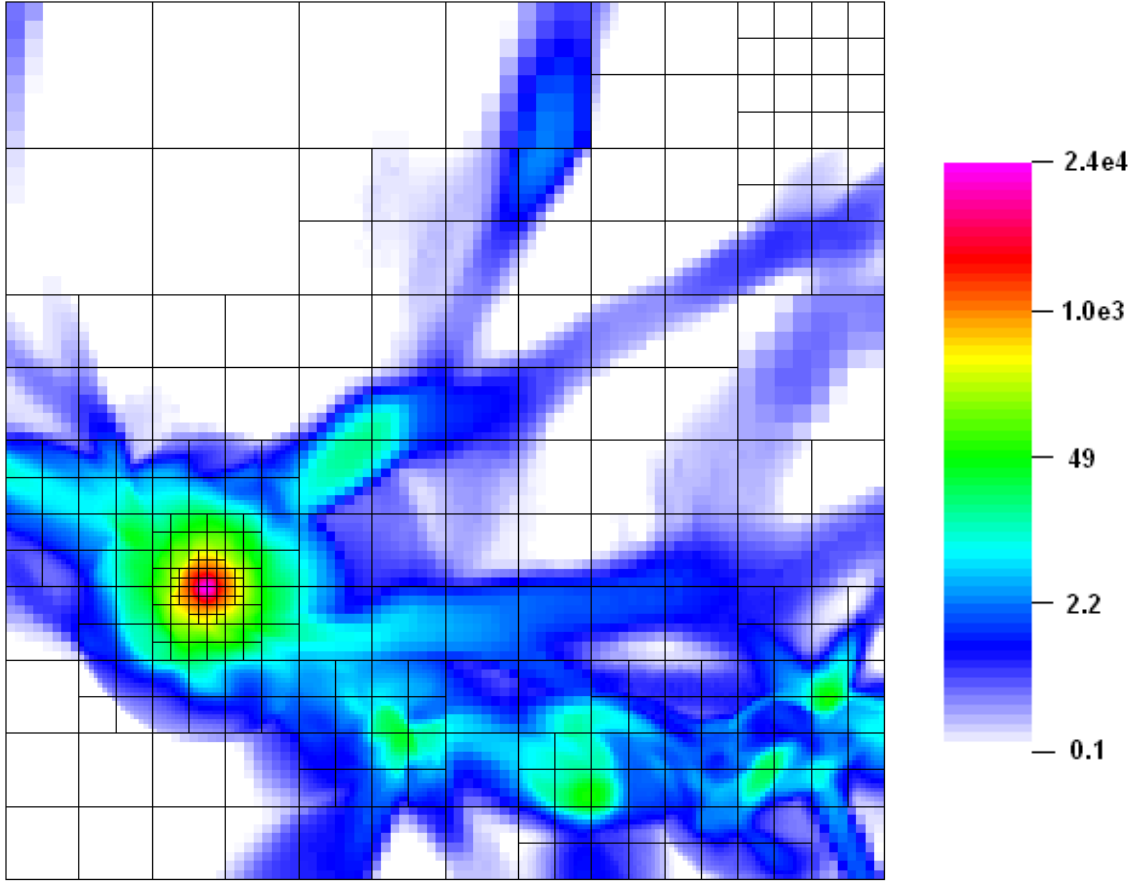}
\caption{Two-dimensional slice of the gas density distribution in logarithm scale. The image is obtained in a purely-baryonic cosmological simulation at $z=0$, and the image size is $18.75\times18.75~h^{-1}$Mpc in a $(100~h^{-1}{\rm Mpc})^3$ simulation box. The refinement map is also shown for illustration, in which each grid represents a single patch. In this snapshot, the densest region is refined to $\ell=4$, giving $4096^3$ effective resolution.}
\label{fig:CosmologyRefinement}
\end{figure}

To test the multi-GPU performance, we perform purely-baryonic cosmological simulations with the same initial condition adopted in the single-GPU case, and measure the performance using $1-32$ GPUs ($N_{{\rm GPU}}=1-32$) in the GraCCA system. In order to have consistent results, the simulation parameters and the refinement criterion for the multi-GPU tests are the same as the ones adopted in the single-GPU test, except that the sizes of the root levels are set to $128^3$ and $256^3$ for the runs with $N_{{\rm GPU}}=1-8$ and $16-32$, respectively. The maximum refinement levels are chosen to be $\ell_{max}=5$ in both cases, giving $4096^3$ and $8192^3$ effective resolutions. Accordingly, the highest comoving spatial resolution is 12 $h^{-1}$Kpc. As an illustration, Figure \ref{fig:CosmologyRefinement} shows a two-dimensional snapshot of the simulation results at $z=0$, in which we plot the density distribution and the corresponding refinement map. Also note that for $N_{{\rm GPU}}=1-16$, we use only one GPU per computing node, whereas for $N_{{\rm GPU}}=32$ we use two GPUs per node in the GraCCA system. Therefore, in all timing tests (except for the case $N_{{\rm GPU}}=1$) the network communications are involved.

\begin{figure}
\centering
\includegraphics[width=15cm]{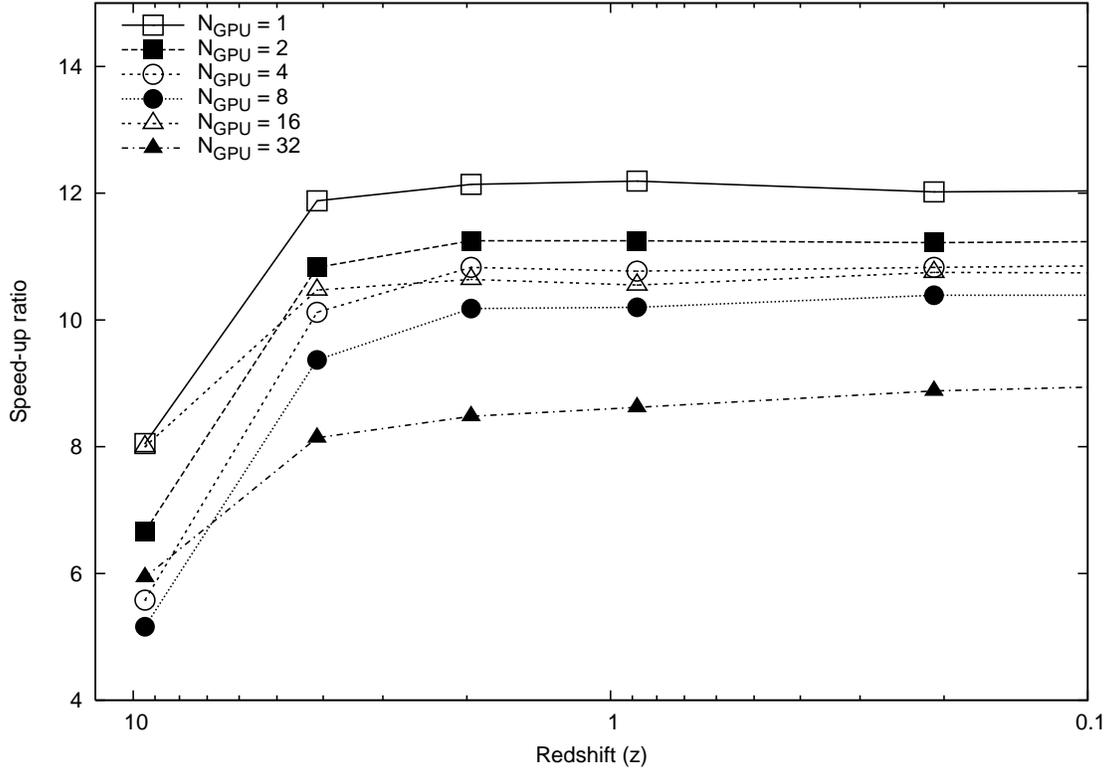}
\caption{Performance speed-up ratios as a function of redshift using $1-32$ GPU(s) in the GraCCA system. We measure the performance in purely-baryonic cosmological simulations. The speed-up ratios are measured by comparing the runs with GPU(s) acceleration to the runs using CPU(s) only. The concurrent execution between CPU and GPU are activated in all tests. The sizes of the root levels are set to $128^3$ and $256^3$ for the runs using $1-8$ and $16-32$ GPU(s), respectively, and the maximum refinement levels are set to 5 in all tests.}
\label{fig:GPUvsCPU_vs_Redshift__NRank1-32}
\end{figure}

Figure \ref{fig:GPUvsCPU_vs_Redshift__NRank1-32} shows the performance speed-up ratio versus redshift for $N_{{\rm GPU}}=1-32$. The ratios are measured by using the same number of CPUs and GPUs in each test. For example, the speed-up ratio for $N_{{\rm GPU}}=16$ is measured by comparing to the timing result using 16 CPUs. The concurrent execution between CPU and GPU are activated in all runs. As expected, for the test runs with the same size of the root level and the same maximum refinement level (and hence the computational workloads are the same), the speed-up factors slightly decrease with the number of GPUs. It is reasonable since the network communication accounts for a larger percentage of total simulation time when using GPUs. Nevertheless, the performance enhancement still exceeds 10.47 after $z\le4.12$ when using 16 GPUs. This result is encouraging since it shows that the total simulation time is still dominated by the computation even with the use of GPUs. Moreover, note that in the GraCCA system different nodes are connected by Gigabit Ethernet, which has a relatively low bandwidth compared to other high-bandwidth interconnections, for example, the InfiniBand and Myrinet. Timing analyses show that the network communication accounts for $8\%-11\%$ of the total simulation time for $N_{{\rm GPU}}=16$.

We also notice that a relatively low speed-up ratio is observed in the case $N_{{\rm GPU}}=32$, in which two GPUs in the same computing node are both activated. This performance drop is mainly caused by the decrease of data transfer bandwidth between one CPU and one GPU. In the GraCCA system, although there are two individual PCI Express buses dedicated for the two GPUs, the total bandwidth actually measured is only marginally above the bandwidth observed in the single-GPU case \citep{Schive08}. In addition, the GeForce 8800 GTX GPUs installed in the GraCCA system do not support the capability of asynchronous memory copies, and hence the data transfer time in the PCI Express bus cannot be overlapped with the GPU kernel executions. However, $8.14-9.24$ speed-up ratios are still demonstrated during $0\le z\le 4.12$ in the $N_{{\rm GPU}}=32$ case. We note that the difficulty with communications via PCI Express bus can potentially be lifted by using the latest motherboard, which supports dual PCI Express 2.0 $\times16$.

\section{CONCLUSIONS AND FUTURE WORKS}

We have presented a novel GPU-accelerated AMR code named \emph{GAMER}, which is dedicated to astrophysical simulations. The AMR implementation is based on constructing a hierarchy of mesh patches with an oct-tree data structure, in which each patch is restricted to contain a fixed number of cells. The hydrodynamic solver is based on a three-dimensional relaxing TVD scheme, which is second-order accurate in both time and space. The gravitational potential is solved by using a multi-level Poisson solver, and the SOR algorithm is adopted to solve the Poisson equation for each mesh patch. The potential error caused by the coarse-fine boundaries is diminished by eliminating the effect of pseudo mass sheets, which are introduced from the discontinuity of resolution across different refinement levels. The computational performance of \emph{GAMER} has been highly improved by an order of magnitude by utilizing the GPU computing power in several parts of the code.

\emph{GAMER} is a parallel code that can be run in a GPU cluster system. The parallelization strategy is based on a rectangular domain decomposition and the concept of using the buffer patches to enclose the computational sub-domain of each CPU/GPU. This method ensures that different patches can be calculated independently and in an arbitrary order, and therefore multiple patches can be computed by GPUs simultaneously. The amount of data transfer has been carefully minimized to avoid the bottleneck of network communication, and a hierarchical search algorithm has been adopted so that the global search for patches adjacent to the boundaries of sub-domain is only necessary at the root level. Currently, a more advanced domain decomposition method using the space-filling curve is being implemented in order to further improve the load balance.

Two GPU kernels have been developed for speeding up the hydrodynamic and Poisson solvers, respectively. For the GPU hydrodynamic solver, the data of each patch group are decomposed into a set of data columns, each of which can be stored in the low-latency shared memory and can be calculated efficiently and simultaneously by GPU. For the GPU Poisson solver, we have developed an elaborate data-exchange algorithm, in which the data of odd and even cells are stored in either the shared memory or the per-thread registers. No potential data transfer between the global memory and the shared memory is required during the SOR iterations, and hence the computational performance is highly improved. In both GPU solvers, in order to reduce the data transfer time in the PCI Express bus, we have utilized the capability of asynchronous memory copies, by which the data transfer time is overlapped with the GPU kernel executions. To reduce the computational overhead associated with the preparation of the ghost-zone data, the eight nearby patches are always grouped into a patch group before sending into GPU. Furthermore, we have exploited the concurrent execution between CPU and GPU, by which the computation times in CPU and GPU can be overlapped with each other.

We have measured the performances of individual GPU solvers in different hardware implementations, including the GraCCA system and the GPU-equipped nodes in CQSE and NCHC. In each performance test, we have compared the performance using GPU acceleration as opposed to the performance using CPU only. Maximum speed-up factors of 15.3 and 24.1 are demonstrated for the hydrodynamic and Poisson solvers, respectively. We have also measured the efficiency of the concurrent execution between CPU and GPU, and the maximum overheads are found to be less than 4.8\% for the hydrodynamic solver and 12.5\% for the Poisson solver.

The accuracy of \emph{GAMER} has been verified by performing several standard test problems, including the shock tube test, Sedov-Taylor blast wave test, and spherical collapse test. The agreement with analytical solutions, the good shock-capturing capability, and the convergence of numerical solutions have been demonstrated. The aspherical scatter is found to be small when solving spherically symmetric problems on the Cartesian grid. We have also shown the significance of the potential correction by comparing the potential solutions obtained by the adaptive-mesh method to those one obtained by the high-resolution uniform-mesh method.

We have measured the performance of the complete \emph{GAMER} code in purely-baryonic cosmological simulations, in which we have adopted effective resolutions $4096^3$ and $8192^3$ when using $1-8$ and $16-32$ GPU(s), respectively. An order of magnitude performance speed-ups have been observed when using $1-16$ GPU(s), which demonstrates the remarkable performance of the code. A relatively low speed-up factor (still exceeding 8) has been observed when using 32 GPUs in the GraCCA system, which is due to the insufficient bandwidth between CPU and GPU when using two GPUs in the same node. This bottleneck should, however, be able to get highly reduced by using the latest motherboard supporting dual PCI Express 2.0 $\times16$ and by utilizing the capability of asynchronous memory copies enabled in the latest GPUs. In these tests, we have also demonstrated the importance of exploiting the concurrent execution between CPU and GPU, by which speed-up factors of $1.4-1.6$ have been measured when compared to the results without the CPU and GPU overlap.

In the present implementation, the performance bottleneck of \emph{GAMER} lies in the computing time of the ghost-zone data for each solver. It is because this calculation is performed by CPU and also because the number of cells in the ghost zone is comparable to that of the interior cells in each patch group. Adopting a higher-order and more arithmetic-intensive algorithm, for example, the third-order PPM scheme, can almost certainly eliminate this bottleneck. Also note that, in the current implementation of \emph{GAMER}, the computation time of the Poisson solver is approximately equal to that of the hydrodynamic solver. This is due to the fact that we apply $4-8$ SOR operations in one step in order to improve the accuracy of gravitational potential. Clearly, investigating a more precise Poisson solver with higher arithmetic intensity can lead to a further performance improvement.

Another approach to further improve performance is to implement more elements of the code in GPU, for example, the interpolation and the oct-tree data structure. However, it will certainly lead to substantial inflexibility of the code. The current GPU implementation in \emph{GAMER} aims at balancing the performance and the flexibility. The GPU kernels are only applied to individual solvers, and the main AMR data structure is still controlled by CPU. Therefore, by simply modifying the GPU kernels, \emph{GAMER} can be applied to different applications and include various physics straightforwardly.

We note that the \emph{GAMER} code can serve as an extremely high-performance and general-purpose tool for astrophysical simulations. Although at present only hydrodynamics and self gravity are included, we believe that we have developed the framework for exploiting the enormous GPU computing power with an astrophysical AMR code. Future works in \emph{GAMER} will focus on including more physics for various simulations. For example, one of the most straightforward extension is to include the dark matter particles and calculate the corresponding density by the standard cloud-in-cell method. Another promising extension is to include the magnetohydrodynamics, which has been successfully implemented in the AMR framework in several works \citep[e.g.,][]{Ziegler05,Fromang06,Collins09}. We can further include several physics in \emph{GAMER}, for example the gas cooling and the feedback mechanism, in order to simulate the galaxy formation. Currently, \emph{GAMER} has been modified to address the detailed halo profile found in the large-scale structure simulation with an extremely light dark matter model \citep{WC09}.

\section{ACKNOWLEDGEMENT}

We would like to thank Shing-Kwong Wong for stimulating discussions. Portions of the simulations in this work were performed in the National Center for High-Performance Computing of Taiwan and the Center for Quantum Science and Engineering of National Taiwan University. We are grateful to Jyh-Shyong Ho and Ting-Wai Chiu for helping conduct simulations in these two institutions. We also want to thank the referee for constructive comments that greatly improve the work. This work is supported in part by the National Science Council of Taiwan under the grants NSC97-2628-M-002-008-MY3 and NSC98-2119-M-002-001, and also by NTU-CQSE (Nos. 98R0066-65, 98R0066-69).


\begin{deluxetable}{lccccccc}
\tabletypesize{\footnotesize}
\tablecolumns{8}
\tablewidth{0pt}
\tablecaption{Detailed timing analysis of the GPU hydrodynamic solver}
\tablehead{ \colhead{Platform}                  &
            \colhead{Downstream}                &
            \colhead{Upstream}                  &
            \colhead{Kernel}                    &
            \colhead{Total\tablenotemark{a}}    &
            \colhead{Total\tablenotemark{b}}    &
            \colhead{CPU\tablenotemark{c}}      &
            \colhead{Speed-up}                  \\
            \colhead{}                          &
            \colhead{($\mu$s)}                  &
            \colhead{($\mu$s)}                  &
            \colhead{($\mu$s)}                  &
            \colhead{($\mu$s)}                  &
            \colhead{($\mu$s)}                  &
            \colhead{($\mu$s)}                  &
            \colhead{}                          }	
\startdata
GraCCA  & 71 & 46 & 252 & 369 & \nodata & 5369 & 14.55  \\
NCHC    & 51 & 64 & 144 & 262 & 177     & 2366 & 13.37  \\
CQSE    & 44 & 29 & 150 & 223 & 177     & 2701 & 15.26  \\
\enddata
\tablenotetext{a}{Timing results with only one stream.}
\tablenotetext{b}{Timing results with four streams.}
\tablenotetext{c}{Timing results using CPU only.}
\tablecomments{Timing results shown here are the execution times per patch group.}
\label{table:TimingHydro}
\end{deluxetable}

\begin{deluxetable}{lccccccc}
\tabletypesize{\footnotesize}
\tablecolumns{8}
\tablewidth{0pt}
\tablecaption{Detailed timing analysis of the GPU Poisson solver}
\tablehead{ \colhead{Platform}                  &
            \colhead{Downstream}                &
            \colhead{Upstream}                  &
            \colhead{Kernel}                    &
            \colhead{Total\tablenotemark{a}}    &
            \colhead{Total\tablenotemark{b}}    &
            \colhead{CPU\tablenotemark{c}}      &
            \colhead{Speed-up}                  \\
            \colhead{}                          &
            \colhead{($\mu$s)}                  &
            \colhead{($\mu$s)}                  &
            \colhead{($\mu$s)}                  &
            \colhead{($\mu$s)}                  &
            \colhead{($\mu$s)}                  &
            \colhead{($\mu$s)}                  &
            \colhead{}                          }
\startdata
GraCCA  & 13 &  8 & 105 & 126 & \nodata & 2034 & 16.14 \\
NCHC    & 10 & 12 &  39 &  61 & 45      & 1055 & 23.44 \\
CQSE    &  8 &  6 &  42 &  56 & 51      & 1230 & 24.12 \\
\enddata
\tablenotetext{a}{Timing results with only one stream.}
\tablenotetext{b}{Timing results with four streams.}
\tablenotetext{c}{Timing results using CPU only.}
\tablecomments{Timing results shown here are the execution times per patch group.}
\label{table:TimingPoisson}
\end{deluxetable}

\begin{deluxetable}{lrrcccccr}
\tabletypesize{\footnotesize}
\tablecolumns{8}
\tablewidth{0pt}
\tablecaption{Detailed timing analysis of the performance of \emph{GAMER}}
\tablehead{ \colhead{Platform}                          &
            \colhead{Hydro.\tablenotemark{a}}           &
            \colhead{Poisson\tablenotemark{b}}          &
            \colhead{Acc.\tablenotemark{c}}             &
            \colhead{TimeStep}                          &
            \colhead{Hydro. Buffer\tablenotemark{d}}    &
            \colhead{Poisson Buffer\tablenotemark{e}}   &
            \colhead{Refinement}                        &
            \colhead{Total}                             \\
            \colhead{}                                  &
            \colhead{(s)}                               &
            \colhead{(s)}                               &
            \colhead{(s)}                               &
            \colhead{(s)}                               &
            \colhead{(s)}                               &
            \colhead{(s)}                               &
            \colhead{(s)}                               &
            \colhead{(s)}                               }
\startdata
GraCCA (GPU-async)  &  11.14 &  12.82 & 1.95 & 0.01 & 0.25 & 0.34 & 0.36 &  26.88   \\
GraCCA (GPU-sync)   &  18.54 &  20.07 & 2.58 & 0.01 & 0.26 & 0.34 & 0.40 &  42.21   \\
GraCCA (CPU)        & 131.51 & 191.01 & 2.81 & 2.26 & 0.26 & 0.34 & 0.40 & 328.60   \\
NCHC (GPU-async)    &   6.58 &   6.51 & 1.35 & 0.00 & 0.16 & 0.28 & 0.27 &  15.20   \\
NCHC (GPU-sync)     &  10.49 &  10.86 & 2.08 & 0.00 & 0.15 & 0.27 & 0.28 &  24.17   \\
NCHC (CPU)          &  58.95 & 101.16 & 1.25 & 0.99 & 0.16 & 0.27 & 0.27 & 163.09   \\
CQSE (GPU-async)    &  10.70 &   9.74 & 1.86 & 0.00 & 0.24 & 0.30 & 0.35 &  23.19   \\
CQSE (GPU-sync)     &  15.59 &  14.51 & 2.42 & 0.00 & 0.24 & 0.29 & 0.36 &  33.41   \\
CQSE (CPU)          &  96.88 & 131.49 & 1.52 & 1.34 & 0.24 & 0.29 & 0.34 & 232.10   \\
\enddata
\tablenotetext{a}{Hydrodynamic solver (including the preparation and closing steps).}
\tablenotetext{b}{Poisson solver (including the preparation and closing steps).}
\tablenotetext{c}{Gravitational acceleration.}
\tablenotetext{d}{Preparing the buffer patches for the hydrodynamic solver.}
\tablenotetext{e}{Preparing the buffer patches for the Poisson solver.}
\tablecomments{Timing results shown here are measured in purely-baryonic cosmological simulations at $z=0.88$. The abbreviations ``async'' and ``sync'' represent the runs with and without the concurrent execution between CPU and GPU, respectively. The computation times of the refinement operation have been divided by 4, since in these runs we reconstruct the refinement map every 4 steps. Also note that 4 sibling relaxation steps are applied in the end of the Poisson solver in these timing tests.}
\label{table:TimingGAMER}
\end{deluxetable}

\clearpage


\clearpage

\end{document}